\documentclass[11pt,a4paper]{article}

\usepackage[utf8]{inputenc}
\usepackage{xcolor}
\definecolor{grey}{rgb}{0.3,0.3,0.3}
\usepackage{amsmath}
\usepackage{amsfonts}
\usepackage{float}
\usepackage{graphicx}
\usepackage{hyperref}
\usepackage[numbers]{natbib}
\usepackage{microtype}
\usepackage[margin=1in]{geometry}
\usepackage{soul}
\usepackage{placeins}

\title{Business Entity Entropy}
\author{Adam McCabe\textsuperscript{\dag} and Matthew H. Chequers\textsuperscript{\dag}\\[6pt]
\textsuperscript{\dag}Convictional Research\\
\textcolor{grey}{\texttt{adam.mccabe@convictional.com, matt.chequers@convictional.com}}}
\date{\today}

\begin{document}

\maketitle

\begin{abstract}
\noindent Organizations generate vast amounts of interconnected content across various platforms. While language models enable sophisticated reasoning for use in business applications, retrieving and contextualizing information from organizational memory remains challenging. We explore this challenge through the lens of entropy, proposing a measure of entity entropy to quantify the distribution of an entity's knowledge across documents as well as a novel generative model inspired by diffusion models in order to provide an explanation for observed behaviours. Empirical analysis on a large-scale enterprise corpus reveals heavy-tailed entropy distributions, a correlation between entity size and entropy, and category-specific entropy patterns. These findings suggest that not all entities are equally retrievable, motivating the need for entity-centric retrieval or pre-processing strategies for a subset of, but not all, entities. We discuss practical implications and theoretical models to guide the design of more efficient knowledge retrieval systems.
\end{abstract}

\section{Motivation and Intuition}
Organizations today generate a continuous stream of content -- documents, meeting transcripts, emails, code repositories, database entries, etc -- accumulating into vast, interconnected knowledge bases. Within this evolving digital environment, organizations face the fundamental challenge of knowledge creation and management \cite{nonaka1994dynamic}. Leaders and decision-makers increasingly rely on natural language processing and large language models (LLMs) to surface relevant insights and inform strategic choices. While powerful language models now enable sophisticated reasoning and summarization through retrieval-augmented generation (RAG) techniques \cite{lewis2020retrieval}, a crucial challenge remains: effectively retrieving and contextualizing the right information from the sprawling organizational memory.

Standard RAG pipelines assume that relevant knowledge can be found by pulling a handful of pertinent documents and feeding them to a language model. This approach is effective when the necessary facts cluster together in a small number of documents. However, not all business entities or topics are neatly confined. Some are indeed ``low-entropy'' entities -- key concepts, individuals, products, or initiatives whose essential information resides in just one or two documents. Retrieving their full context is straightforward. Yet, other entities display ``high-entropy'' characteristics -- they are diffusely mentioned across dozens or even hundreds of documents, with no single source containing a majority of their facts. For such entities, simple document retrieval hits a scalability wall: either the system omits critical information or must include a large swath of potentially irrelevant material, leading to inefficiency and incomplete answers.

This pattern hints at an underlying structure in how organizational knowledge distributes itself. Some entities remain tightly bound to a few canonical sources; others fragment unpredictably across the knowledge base, defying simple retrieval strategies. Drawing on information theory, we conceptualize this dispersion as an ``entropic'' property of the entity's knowledge footprint. Building on foundational work in information theory \cite{shannon1948mathematical,cover1991elements}, we propose that the varying degrees of entity ``spread" can be rigorously measured and analyzed.

Why entropy? In statistical physics, entropy measures the number of possible `micro-state' arrangements that give rise to the same `macro-state' observation. We can apply this concept to our context in that many different arrangements of facts across documents can lead to the same overall knowledge about an entity. When facts about an entity are concentrated in a single source, there are few possible arrangements, leading to low entropy. However, when facts are evenly spread across hundreds of documents, there are many possible arrangements, resulting in high entropy. This physical interpretation maps elegantly to information theory, where entropy measures uncertainty about which message -- or in our case, which document -- will appear next. Both interpretations are equivalent \cite{shannon1948mathematical}, and capture the same fundamental idea: when information is highly dispersed, the system faces greater uncertainty in knowing which documents to retrieve or how to combine them efficiently. For the bulk of this paper, we will use the information theory conceptualization, but at times may draw on intuition from the physical interpretation.

This entropic model of context distribution opens several avenues for both practical system design and theoretical inquiry. Theoretically we find that entropy can be used to understand a generative model describing a feedback mechanism helping to grow entropy early on based on initial importance, while tapering off as entropy saturates to the theoretical maximum (Section~\ref{sec:generative_models}). From a practical standpoint, quantifying entropy allows architects of AI decision-support platforms to distinguish between entities that can be handled by simple retrieval methods and those that require more sophisticated, entity-centric pre-processing or more sophisticated retrieval strategies (Section~\ref{sec:practical_use_cases}). From a theoretical perspective, understanding why some entities become ``entropic" and others remain ``low-entropy" could shed light on the underlying generative processes in organizational knowledge formation, providing practical identification algorithms for organizations.

Moreover, introducing an entropy framework sets the stage for deeper questions about how knowledge evolves over time, how it compares across different organizations or domains, and whether certain types of entities inherently resist stable concentration (Section~\ref{sec:empirical_results}). By quantifying these distributions, we not only provide a new lens to analyze organizational memory but also lay the groundwork for a richer theoretical understanding of how human-created documents shape -- and are shaped by -- contextual complexity.

The application of entropy to named entities is not novel. Early work by \cite{chieu2003named} demonstrated the effectiveness of maximum entropy frameworks for named entity recognition (NER), using entropy to model the uncertainty in entity classification across different contexts. However, our work departs from this traditional use of entropy in NER tasks. Rather than applying entropy to improve entity detection and classification, we propose a novel entropy measure that quantifies how entity knowledge distributes itself across an organization's document corpus.

This work aims to bridge the gap between the engineering pragmatics of RAG and a richer, more mathematically grounded understanding of enterprise knowledge distribution. In the following sections, we will formalize intuitions related to entropy in Section~\ref{sec:defining_entropic_measures}, present empirical findings in Section~\ref{sec:empirical_results}, propose theoretical models that explain the observed patterns in Section~\ref{sec:generative_models}, and discuss practical applications in Section~\ref{sec:practical_use_cases}. Finally, in Section~\ref{sec:conclusion} we highlight our conclusions. Throughout this paper we will refer to the Appendix in Section~\ref{sec:appendix} for supporting material.

\FloatBarrier
\section{Defining the Entropic Measures} \label{sec:defining_entropic_measures}

To systematically capture the intuitive notion of ``contextual sprawl'', we introduce a formal measure of entropy inspired by information theory. Just as Shannon entropy quantifies the uncertainty in a probability distribution over possible messages \cite{shannon1948mathematical}, we apply a similar framework to measure how evenly facts about a given entity are distributed across a set of documents.

\subsection{Setting and Notation}

Consider an organization's corpus of documents, $\mathcal{D} = \{d_1, d_2, \ldots, d_N\}$, where each $d_i$ is a text artifact (e.g., a meeting transcript, a report, a GitHub issue, etc, or chunks of such) produced or archived by the business. Within this corpus, we focus on a particular physical or conceptual entity \cite{ratinov2009design} $E$, which could be a product, a key individual, an initiative, or a concept relevant to the business. Our goal is to understand how the knowledge associated with $E$ is distributed across $\mathcal{D}$.

For each document $d \in \mathcal{D}$, let $f_{E}(d)$ represent the number of distinct facts about the entity $E$ that appear in $d$. Facts represent single pieces of information about an entity, for example ``Person X is the CEO of Company Y''. These facts might be extracted through NER and entity-level fact extraction pipeline (see Appendix, Section~\ref{appendix:fact_extraction_pipeline}), or any suitable information extraction approach. We assume that $\sum_{d \in \mathcal{D}} f_{E}(d) > 0$, meaning that $E$ is mentioned in at least one document.

\subsection{Probability Distributions Over Documents}

To apply the concept of entropy, we must first define a probability distribution. Given entity $E$, we define the probability of a document $d$ containing information about $E$ as follows

\begin{equation} \label{eq:probability}
p_{E}(d) = \frac{f_{E}(d)}{\sum\limits_{d' \in \mathcal{D}} f_{E}(d')}.
\end{equation}

\noindent In other words, Equation \ref{eq:probability} represents the fraction of $E$'s known facts found in document $d$. By construction, $\sum_{d \in \mathcal{D}} p_{E}(d) = 1$. This simple model overlooks fundamental issues such as fact duplication, noise, or the quality of fact extraction (more on these in Section \ref{sec:limitations}). However, it provides a useful starting point for quantifying the distribution of entity knowledge across documents and suffices for the purposes of this paper.

This probability distribution reflects the ``concentration'' of information about $E$. If a single document dominates and contains most of $E$'s facts, then $p_{E}(d)$ will be heavily skewed toward that document. Conversely, if facts are spread evenly across many documents, the distribution $p_{E}(d)$ will approach uniformity.

\subsection{Shannon Entropy}

Given the distribution in Equation \ref{eq:probability}, we define the entropy of an entity $E$\footnote{It is important to note that technically this is only an estimate on an upper bound for the entity's entropy. Duplication of facts as well as fact extraction pipeline accuracy both may lead to reduction in entropy.} as

\begin{equation} \label{eq:entropy}
H(E) = -\sum_{d \in \mathcal{D}} p_{E}(d) \log p_{E}(d).
\end{equation}

Equation \ref{eq:entropy} is the Shannon entropy of the distribution $p_{E}$, building on established information theory foundations \cite{cover1991elements}. It quantifies how ``uncertain'' one would be if asked to guess which document contains a randomly chosen fact about $E$. A reminder on the basic key properties and what they mean in our context include:

\begin{itemize}
\item \textbf{Minimum Entropy (0):} If all of $E$'s facts are in a single document, say $f_{E}(d_{\text{max}}) = \sum_{d'} f_{E}(d')$, then $p_{E}(d_{\text{max}})=1$ and $p_{E}(d) = 0$ for $d \neq d_{\text{max}}$. The entropy is
\[
H(E) = -(1 \cdot \log 1 + 0 + \cdots + 0) = 0.
\]
A zero entropy entity is thus one with perfectly localized knowledge, easy to retrieve via modern search techniques.

\item \textbf{High Entropy:} If $E$ appears uniformly across $M$ distinct documents, with each document holding $\frac{1}{M}$ of the facts, then
\[
H(E) = -\sum_{i=1}^{M} \frac{1}{M} \log \frac{1}{M} = \log M.
\]
In this scenario the entropy increases with the number of documents that evenly share $E$'s facts, creating scaling challenges for modern LLM's context windows and issues with holistically ranking documents on more than semantic similarities (e.g., causal relationships).

\item \textbf{Sensitivity to Distribution Shape:} Entropy is not just about the number of documents; it also depends on how facts are distributed across documents. It is for this reason that we should not simply look at counts of documents with mentions of $E$. For example, if one document holds 50\% of the facts and four others hold 12.5\% each, the distribution is less uniform than a situation where all five documents hold exactly 20\%. The entropy metric will reflect these differences.
\end{itemize}

\subsection{Choice of Logarithm Base and Units}

The entropy measure is often independent of the base of the logarithm, differing only by a constant factor. By default, information theory uses base 2, measuring entropy in bits. Using natural logarithms ($\ln$) measures entropy in nats. The choice is largely a matter of convention and does not affect the relative comparisons between entities; in our case we have chosen base 2 to remain consistent with convention in information theory.

\subsection{Practical Retrieval Constraints and ``Capacity''}

While Shannon's notion of channel capacity refers formally to the maximum reliable information rate in a noisy communication channel \cite{shannon1948mathematical}, we can loosely borrow that language to illustrate a core tension in RAG. Modern LLMs still have bounded context windows, meaning only a limited number of documents (or chunks) can be fed as context in a single query. If an entity $E$ has very high entropy, its associated facts are widely scattered. Covering them in a single pass can force us to retrieve a large set of documents -- potentially exhausting context limits causing additional passes and introducing noise.

\paragraph{Context Budget Analogy.}
Suppose we have $N$ documents available for retrieval. If an entity’s facts are nearly all concentrated in one or two of those documents (i.e., low-entropy), it is straightforward to select them within our limited “context budget.” Conversely, high-entropy entities require multiple documents for coverage, crowding the context window with potentially duplicative or tangential information. In this sense, high entropy reduces the effective ``capacity'' to retrieve relevant facts in a single pass.

\paragraph{Implication: Pre-Process High-Entropy Entities.}
A practical consequence of this observation is that \emph{identifying} high-entropy entities helps us \emph{pre-process} them via summarization or fact consolidation. By assembling a structured overview for these complex entities, we effectively reduce their entropy before engaging the LLM. This frees up space within the context window to include additional supporting information or to handle multiple entities in the same query. As demonstrated in Section~\ref{sec:practical_use_cases}, systematically lowering the entropy of critical entities can significantly improve retrieval performance and reduce the noise that arises when many documents must be simultaneously retrieved.

\paragraph{Note on the Analogy.}
Although we use the term ``capacity'' to illustrate why high-entropy entities are costly to handle, it should be viewed as a heuristic rather than a rigorous application of Shannon’s capacity theorems. We do not claim a formal one-to-one mapping between entity entropy and channel coding mechanisms; rather, we draw on the concept to highlight that a limited retrieval “budget” competes with the need to represent high-entropy knowledge.

\subsection{Experimental Limitations} \label{sec:limitations}

\begin{itemize}
\item \textbf{Normalization:} Some entities may appear in only a small number of documents. While entropy can still be calculated, comparing low-frequency entities to high-frequency ones might require considering sample size or introducing thresholds for minimum mention counts. In our case we do not perform normalization given the relatively small nature of our underlying document corpus and our ability to parse each document individually without scale concerns.

\item \textbf{Chunking and Document Granularity:} In practice, documents may be split into smaller chunks for indexing or retrieval. The entropy measure can be applied at any chosen level of granularity. However, more aggressive chunking can artificially increase entropy. For our purposes we work with complete documents, only chunked when context window limits become a concern for our extraction pipeline. This allows us to maintain a consistent document-level entropy measure without introducing addtional parameter choices in the form of the chunk size and strategy.

\item \textbf{Fact Extraction Quality:} The accuracy of the underlying fact extraction process influences $p_{E}(d)$. Noisy extraction can inflate entropy or bias distributions. Thus, entropy is always conditioned on the quality of the input pipeline. We use state-of-the-art foundation languge models for our fact extraction pipeline, but acknowledge that there are limitations to this approach.

\item \textbf{Entity and Fact Overlap:} Entities may share facts and/or documents may contain duplicate facts leading to overlap in their entropy distributions. This overlap can complicate retrieval strategies and may require de-duplication or more sophisticated entity linking techniques. We do not address this issue in this paper, but it is an important consideration for future work. We will briefly discuss this topic further in Section~\ref{sec:document_overlap}.

\item \textbf{Bias in Empirical Data:} Given that our empirical analysis is based on the corpora of a relatively small organization over a short (12 month) period, the results may not generalize to larger or more diverse datasets. However, the patterns we observe are likely to hold across a wide range of organizations and document types. We will discuss this further in Section~\ref{sec:empirical_results} and note this as a potential area for improvement in future work.
\end{itemize}

\subsection{The Temporal Nature of Entity Entropy}

The entropy of an entity is not static but evolves over time as new documents are created and old ones are updated, deleted or simply become out of date. This temporal aspect is crucial for understanding how knowledge is accumulated and distributed across an organization. As new facts are added to the entity, the entropy will increase, while the staleness of old facts will add noise reflecting the uncertainty in where the new facts will appear. Conversely, as facts are removed or consolidated, the entropy will decrease, indicating a more concentrated knowledge state. We touch on this more in Section \ref{sec:generative_models} based on observations in Section \ref{sec:empirical_results}.

\FloatBarrier
\section{Empirical Results} \label{sec:empirical_results}

In this section, we present empirical results from a large-scale analysis of entities extracted from a startup's content corpus (limited to 2024). Our dataset comprises thousands of documents sourced from diverse repositories -- ranging from internal meeting notes and project documentation, to GitHub issues and external reports. From this corpus, we identified and extracted facts about entities of various categories (e.g., products, individuals, teams, and initiatives) and computed their entropies as defined in Section \ref{sec:defining_entropic_measures}.

\subsection{Dataset and Extraction Methodology}

We processed approximately 2,400 documents, extracting entities and their associated facts using a pipeline that combined LLMs and learned embeddings for named entity recognition, fact extraction and de-duplication support. Entities included tangible resources (e.g., a software tool), intangible concepts (a market segment), and organizational constructs (teams, roles, initiatives). Each fact was a piece of context or detail relevant to the entity, often spanning product descriptions, timelines, relationships, or decision rationales.

Our analysis identified a total of 3,281 unique entities across the corpus. After extraction, each entity's facts were aggregated and linked back to their source documents. We then computed $p_{E}(d)$ and $H(E)$ for each entity using the formulations in Section \ref{sec:defining_entropic_measures}.

\subsection{Distribution of Entropies}

\begin{figure}[H]
\centering
\includegraphics[width=0.8\textwidth]{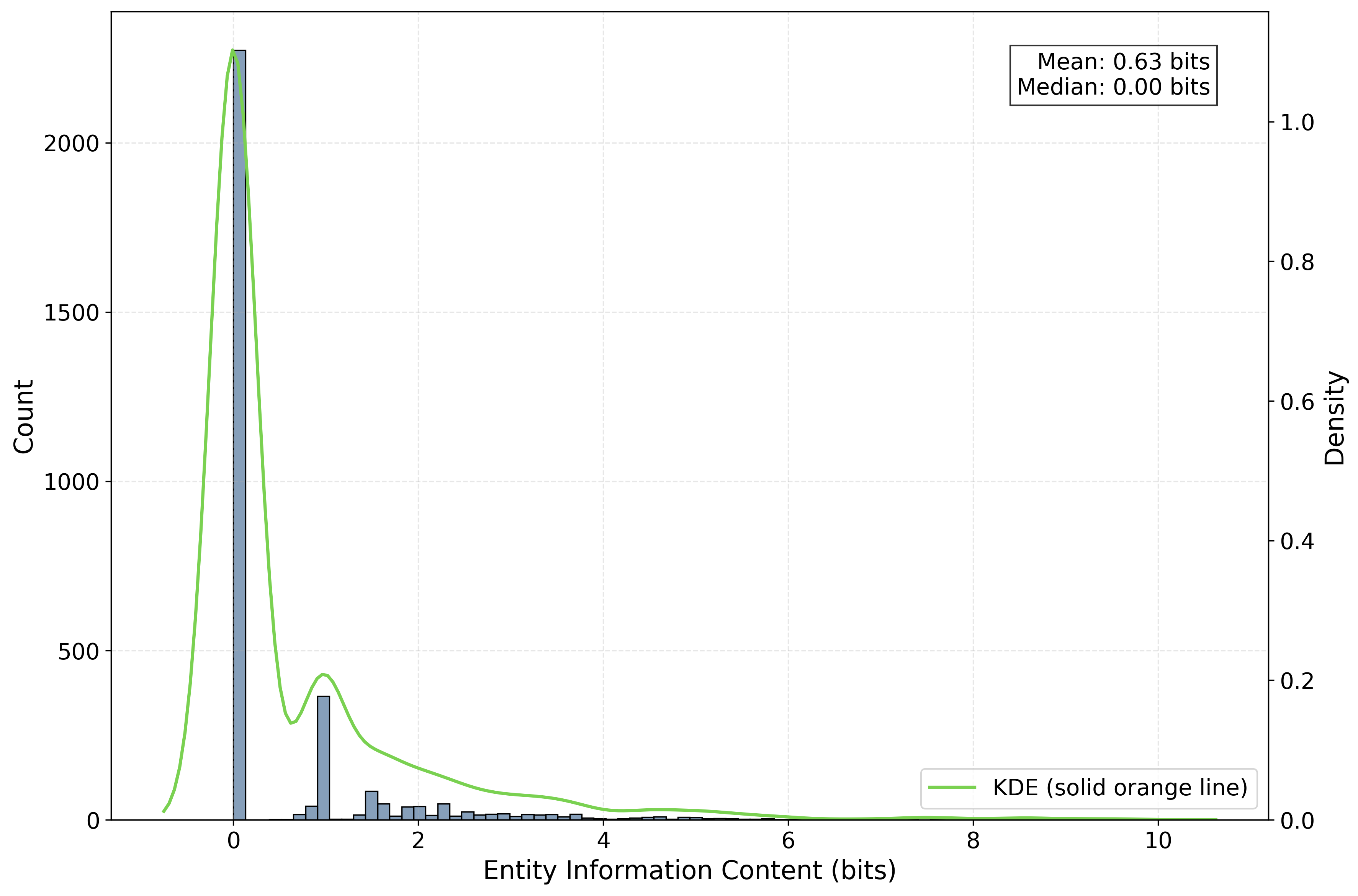}
\caption{Distribution of entity entropies across the corpus, showing a right-skewed distribution with a long tail extending into higher entropy regions. As intuition would suggest it was these entities which were central to how the business operates and included entities such as the company itself, its core product offering, the core repository in which code lived and founding team members. For ease of interpretation we have visualized the distribution (solid green line) using kernel density estimation.}\label{fig:entropy_dist}
\end{figure}

The distribution of entropy across extracted entities is shown in Figure \ref{fig:entropy_dist}. The distribution had a mean and median of 0.63 and 0.00 bits, respectively. For reference, given our document corpus size, the maximum entropy an entity could have is $\approx$ 11.2 bits. While a majority of entities had low to moderate entropy -- indicating that their knowledge was largely concentrated in a few canonical documents -- a small minority exhibited extremely high entropy, with facts scattered roughly evenly across tens or even hundreds of documents. These high-entropy entities were central to how the business operates and included entities such as the company itself, its core product offering, the core repository in which code lived and founding team members. The process driving the evolution of entity entropy is explored further in Section \ref{sec:generative_models}, however, the exogenous factors that can lead to sudden bursts of entropy are not considered in this model.

The heavy-tailed nature of the distribution suggests that a small number of entities are responsible for a disproportionate amount of the corpus's entropy. This pattern is reminiscent of a power-law or Pareto-like behaviour \cite{newman2005power}: a handful of large entities tended to be the most entropic. Smaller entities, in terms of total facts, often had their knowledge captured in just one or two documents, resulting in low entropy. These high-entropy entities are likely to be critical to the organization's operations, reflecting the complex, interconnected nature of modern business knowledge. These are the entities for which pre-processing is often necessary. By maintaing a summarized or targeted store for said entity can in-effect compress it's raw information, greatly reducing its entropy and making it more reliable in RAG pipelines. In contrast, the majority of entities have low entropy, indicating that their knowledge is more localized and easily parsed by both human users and LLMs.

\subsection{Relationship Between Entity Size and Entropy}

\begin{figure}[H]
\centering
\includegraphics[width=0.8\textwidth]{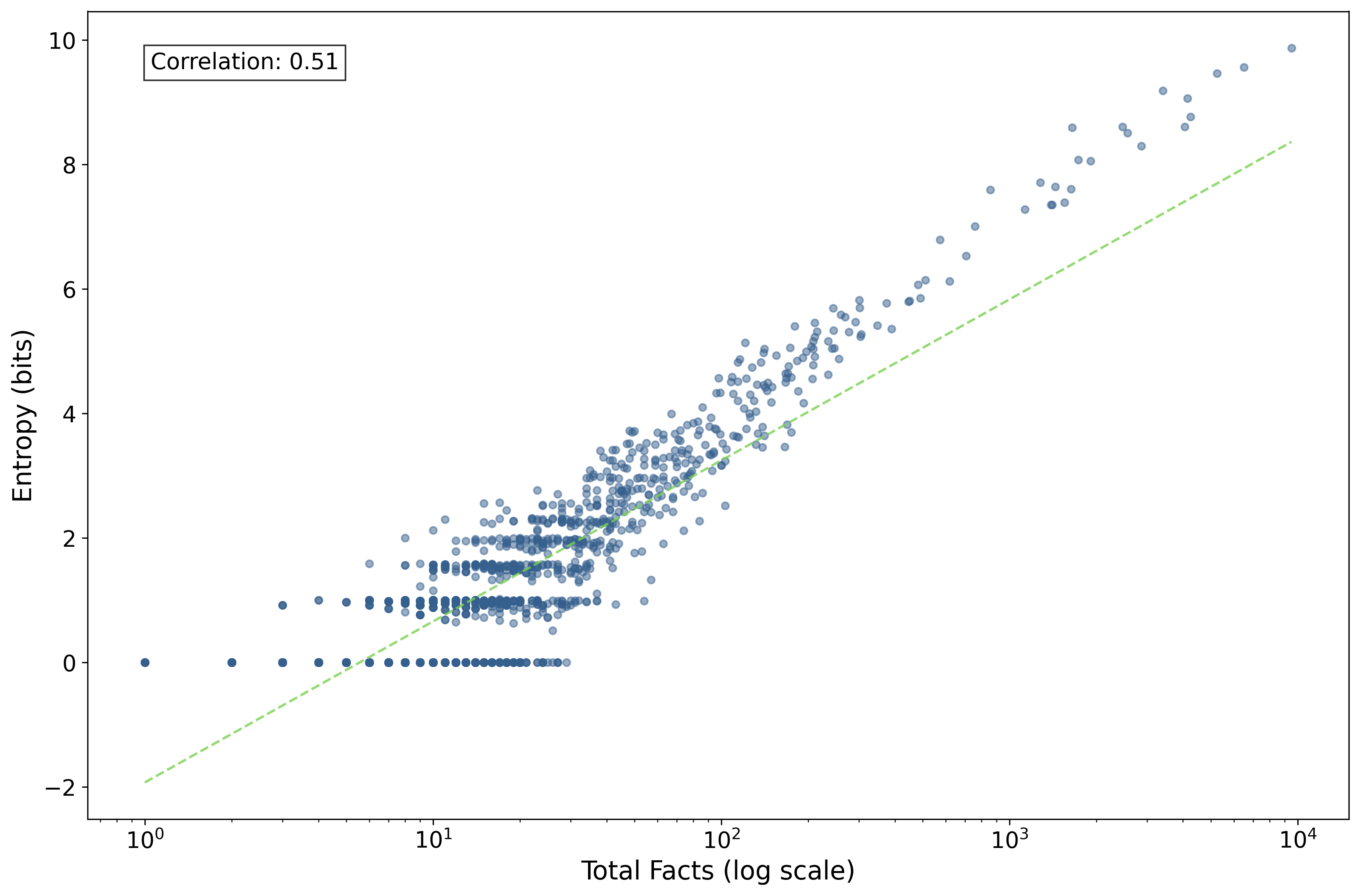}
\caption{Scatter plot showing the relationship between entity size (total facts) and entropy. An example linear relationship is depicted by the dashed line. Note the non-linear relationship between size in facts and an entity's entropy.}\label{fig:size_entropy}
\end{figure}

In Figure~\ref{fig:size_entropy}, we analyzed how the number of total facts associated with an entity (its ``size'') correlated with its entropy. As expected, we found a positive relationship between size and entropy: larger entities tended to have higher entropy, reflecting the challenge of capturing and integrating knowledge about complex, multifaceted entities. This relationship is intuitive: as an entity grows in size or importance to the business, its facts are more likely to be distributed across multiple documents, leading to higher entropy.

As shown in Figure~\ref{fig:size_entropy}, the relationship is not quite linear showing that large entities in terms of facts become more entropic at an increasing rate. This suggests that the complexity of an entity grows faster than its size, reinforcing the challenges of capturing and integrating knowledge about large, critical entities.

\subsection{Relationship Between Entity Category and Entropy}

Our analysis covered several distinct entity types to ensure we only extracted business-relevant entities from the document corpus. Table~\ref{tab:entity-types} shows the entropy statistics for each entity type.

By examining entropy distributions within each category, we found that certain types of entities tend to have lower entropy. While some categories are likely to see similar distributions across all business types, others are much more dependent on the vertical in which the business operates. Our company for example sees low entropy in the `LOCATION' category as we are fully remote, while a real estate company, say, would see high entropy in this category.

\begin{table}[H]
\centering
\caption{Entity Type Statistics}\label{tab:entity-types}
\begin{tabular}{lrrrr}
\hline
Type & Mean Entropy & Median Entropy & Std Dev & Count \\
\hline
COMPANY & 0.69 & 0.00 & 1.29 & 158 \\
COMPETITOR & 0.42 & 0.00 & 0.69 & 82 \\
DEPARTMENT & 0.76 & 0.00 & 1.23 & 39 \\
EXTERNAL REGULATION & 0.42 & 0.00 & 1.03 & 6 \\
INITIATIVE/PROJECT & 0.40 & 0.00 & 0.88 & 735 \\
INTERNAL POLICY & 0.68 & 0.00 & 1.77 & 45 \\
LOCATION & 0.37 & 0.00 & 0.70 & 15 \\
MARKET & 0.42 & 0.00 & 0.79 & 31 \\
PARTNER & 0.77 & 0.00 & 1.11 & 27 \\
PERSON & 1.25 & 0.00 & 2.08 & 253 \\
PROCESS & 0.31 & 0.00 & 0.74 & 406 \\
PRODUCT & 0.57 & 0.00 & 1.28 & 318 \\
ROLE & 0.50 & 0.00 & 0.92 & 47 \\
SYSTEM/TOOL & 0.77 & 0.00 & 1.37 & 972 \\
TEAM & 0.69 & 0.00 & 1.01 & 55 \\
VENDOR & 0.94 & 0.00 & 1.45 & 89 \\
\hline
\end{tabular}
\end{table}

\subsection{Cumulative Coverage and Document Concentration}

Another informative perspective is to consider how many documents are needed to cover a given proportion of an entity's facts. For instance, define a coverage threshold (e.g., 95\%) and measure how many documents are required to reach that threshold. Entities with low entropy typically reach 95\% coverage with just a few documents.

\begin{figure}[H]
\centering
\includegraphics[width=0.8\textwidth]{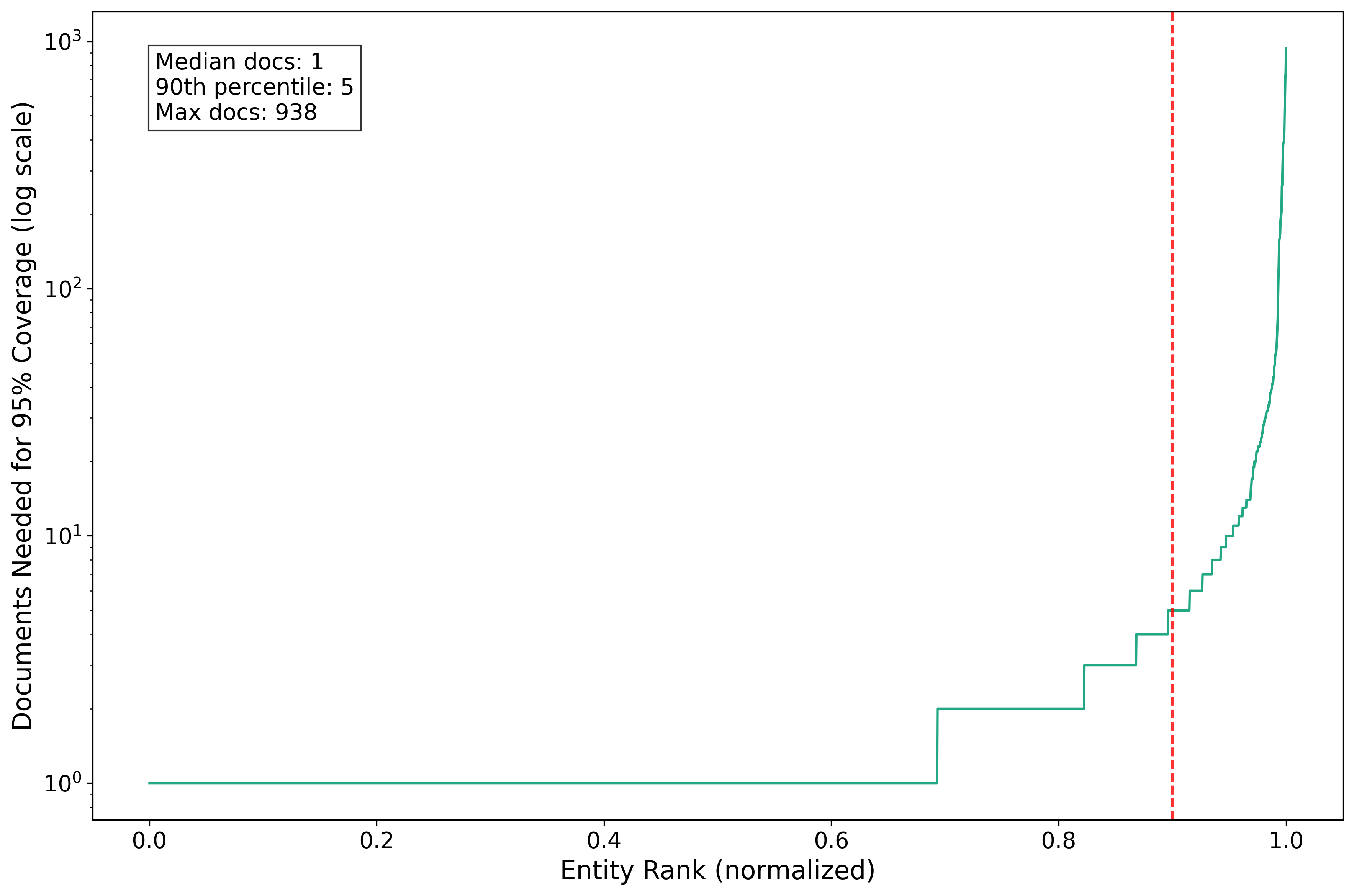}
\caption{Rank-ordered plot of entities by the number of documents needed for 95\% coverage, demonstrating the heavy-tailed nature of document requirements across entities. Vertical dashed red line inserted at the 90th percentile entity.}\label{fig:coverage}
\end{figure}

As we can see in Figure~\ref{fig:coverage}, the distribution of document requirements for 95\% coverage is heavy-tailed, with over 95\% of entities requiring fewer than 10 documents to reach the threshold. However, a small number of entities require dozens or even hundreds of documents to achieve the same level of coverage.

\subsection{Document Overlap}
\label{sec:document_overlap}

When querying for multiple entities, it may be beneficial to retrieve documents that contain information about all entities at once. However, in practice this is not often the case unless specific pre-processing has been performed. This implies that at query time we may need to retrieve documents for each entity individually, leading to increased retrieval times and potential noise in the retrieval process. To understand the impact of this we examined the graph formed from the set of entities, with edges defined if and only if two entities share a document.

\begin{figure}[H]
    \centering
    \begin{minipage}{0.45\textwidth}
        \centering
        \includegraphics[width=\textwidth]{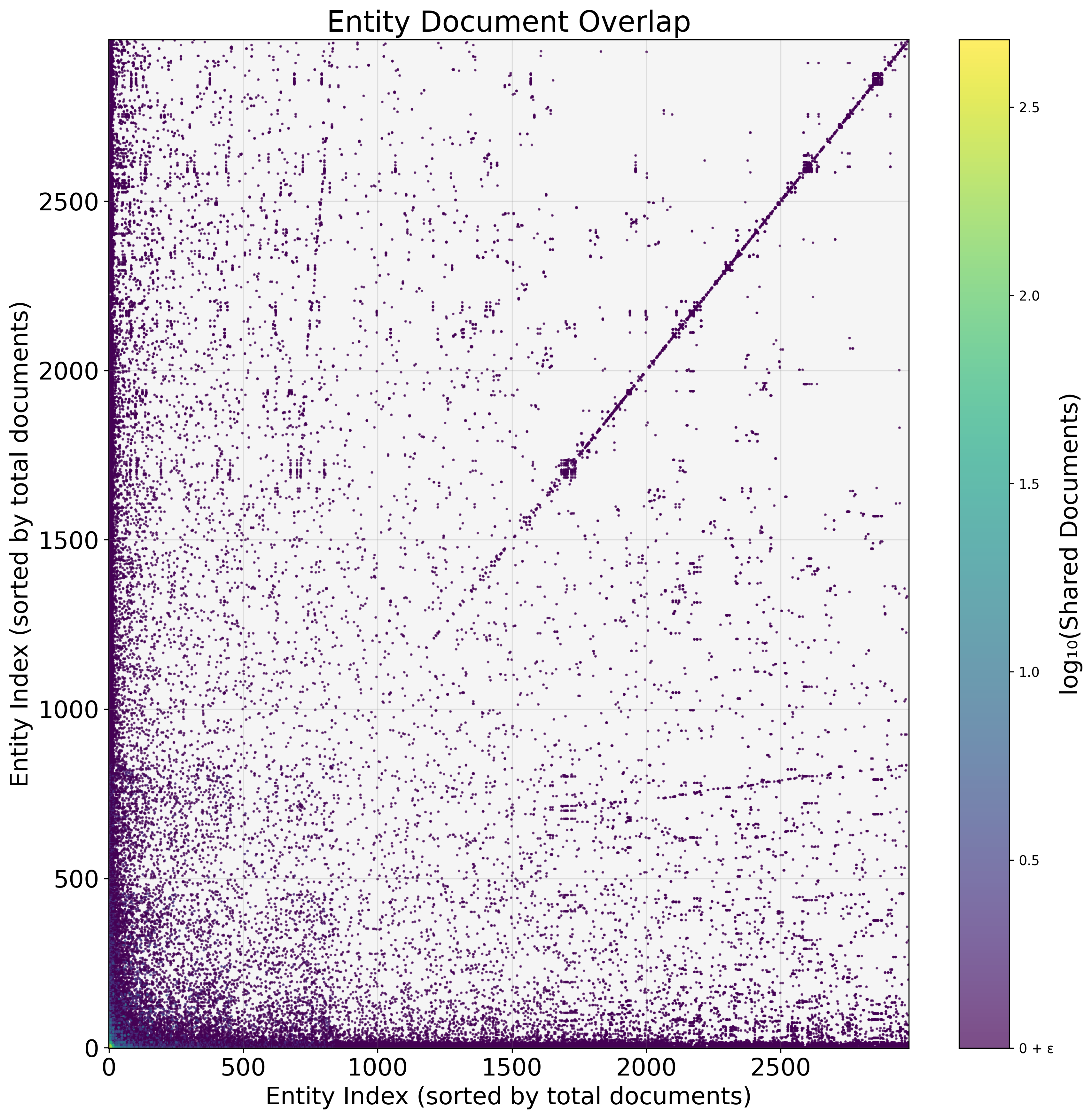}
        \label{fig:all_entity_overlap}
    \end{minipage}
    \hfill
    \begin{minipage}{0.45\textwidth}
        \centering
        \includegraphics[width=\textwidth]{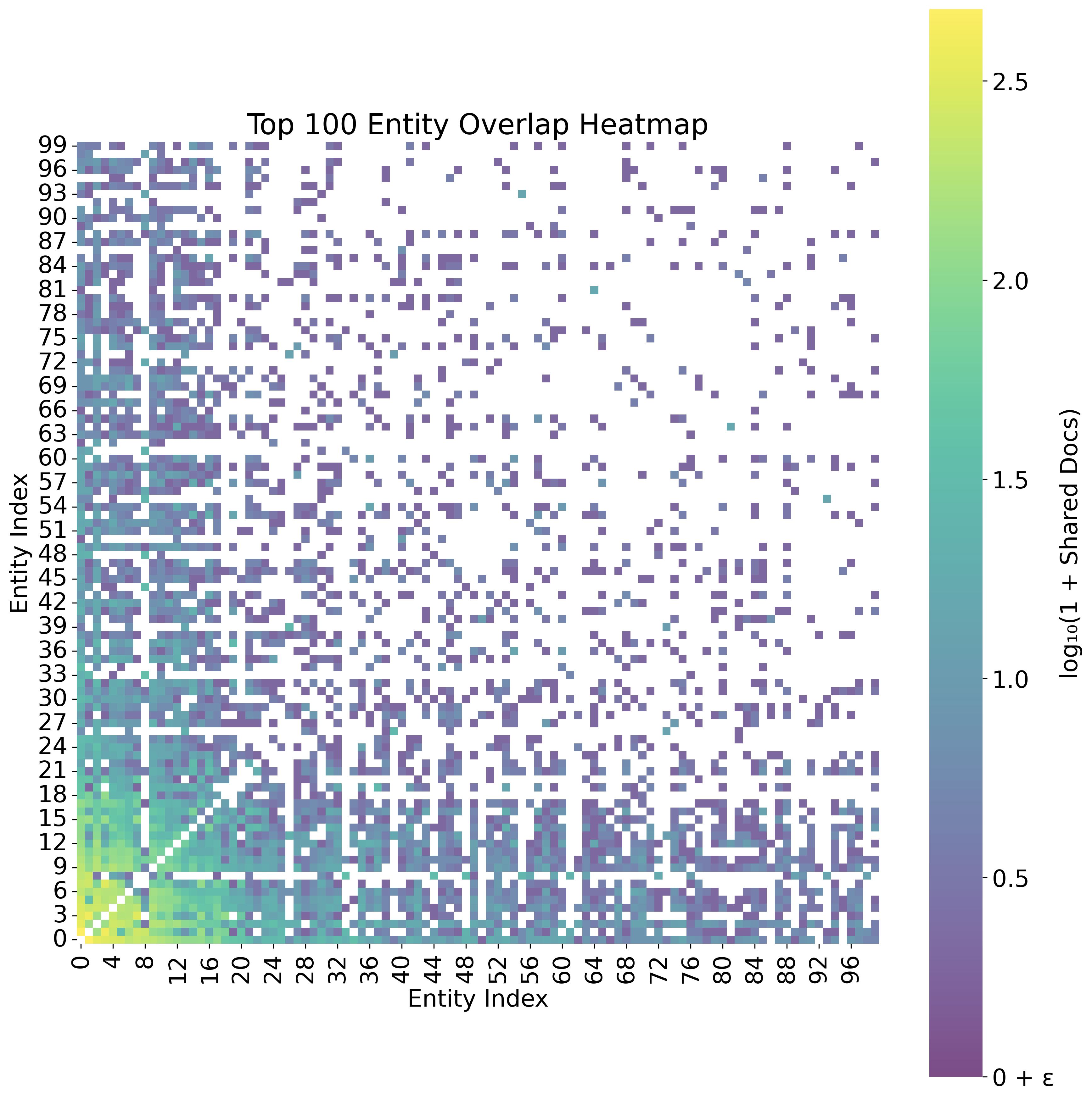}
        \label{fig:top_100_entity_overlap}
    \end{minipage}
    \caption{Heatmap visualizing the adjacency matrix of the connectivity graph for underlying entities. Connectivity is defined as having a shared document, edge weighting according to the number of shared documents. The left plot shows all entities ordered decreasing by total associated documents (points exagerated for readability), while the right plot zooms in on the bottom left corner to show only the top 100 entities. Most points are dark, indiciating low connectivity, the exception being the most entropic entities (as expected). Note, an empty cell (white) represents no shared documents.}
    \label{fig:overlap}
\end{figure}

In examining this graph, represented by an adjacency matrix in Figure~\ref{fig:overlap}, we found that generally most entities do not share documents; however the core entities on the other hand are highly connected themselves and often connected to lower entropy entities. We observed a total network connectivity\footnote{Here \textit{network connectivity} is the proportion of how connected a given graph is compared with a fully connected graph of the same node count.} of 1.2\% across all entities, that jumps to approximately 72\% when considering only the top 100 entities by total associated documents. This suggests that most entities have unique document footprints, while a small number of high entropy entities share a significant number of documents. This overlap can actually benefit real world queries in which most relate to several entities at once, but can also lead to noise in the retrieval process.

\subsection{Exploring the Temporal Nature of Entity Entropy}

\begin{figure}[H]
\centering
\includegraphics[width=0.8\textwidth]{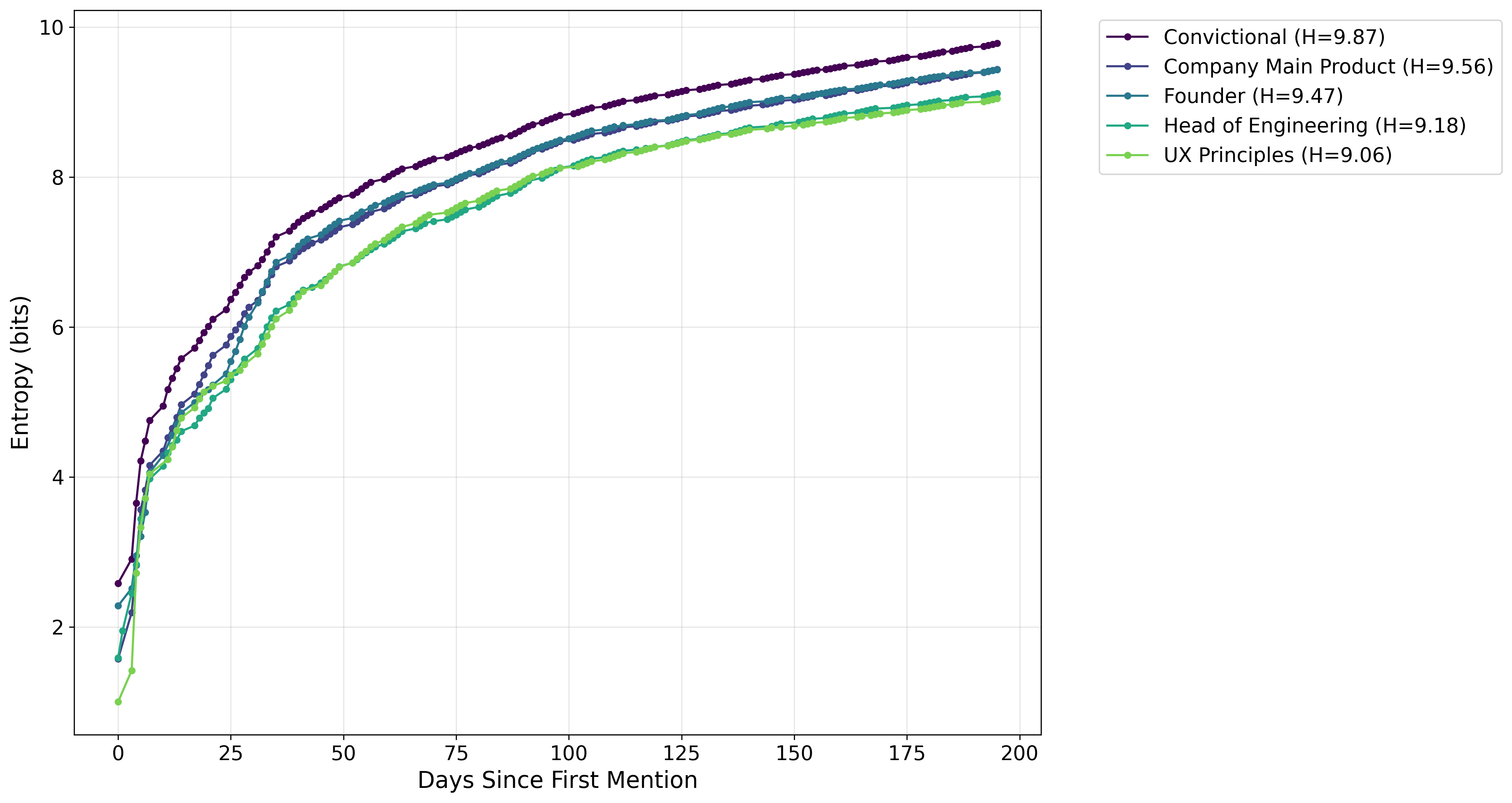}
\caption{Temporal evolution of entropy for the top 5 most entropic entities in the dataset. The behaviour of these largest entities is very similar and likely due to the high overlap in shared documents (see Section \ref{sec:document_overlap})}\label{fig:temporal}
\end{figure}

The entropy of an entity is not static. As new documents are added, updated or deleted the entropy of the given entity varies. We calculate the entropy over time (days from first mention) for a subset of entities and find that the rate of entropy increase is not one-size fits all. In Figure~\ref{fig:temporal} we see how entropy grows for the (currently) top 5 most entropic entities in our dataset. The highest entropy entities continue to grow and represent foundational business entities: the company itself, its core product offering, and the founder. The fifth largest entity, shown in teal, is ``UX Principles'' which once introduced were quickly documented across the corpus but became static except for a few updates.

\begin{figure}[H]
\centering
\includegraphics[width=0.8\textwidth]{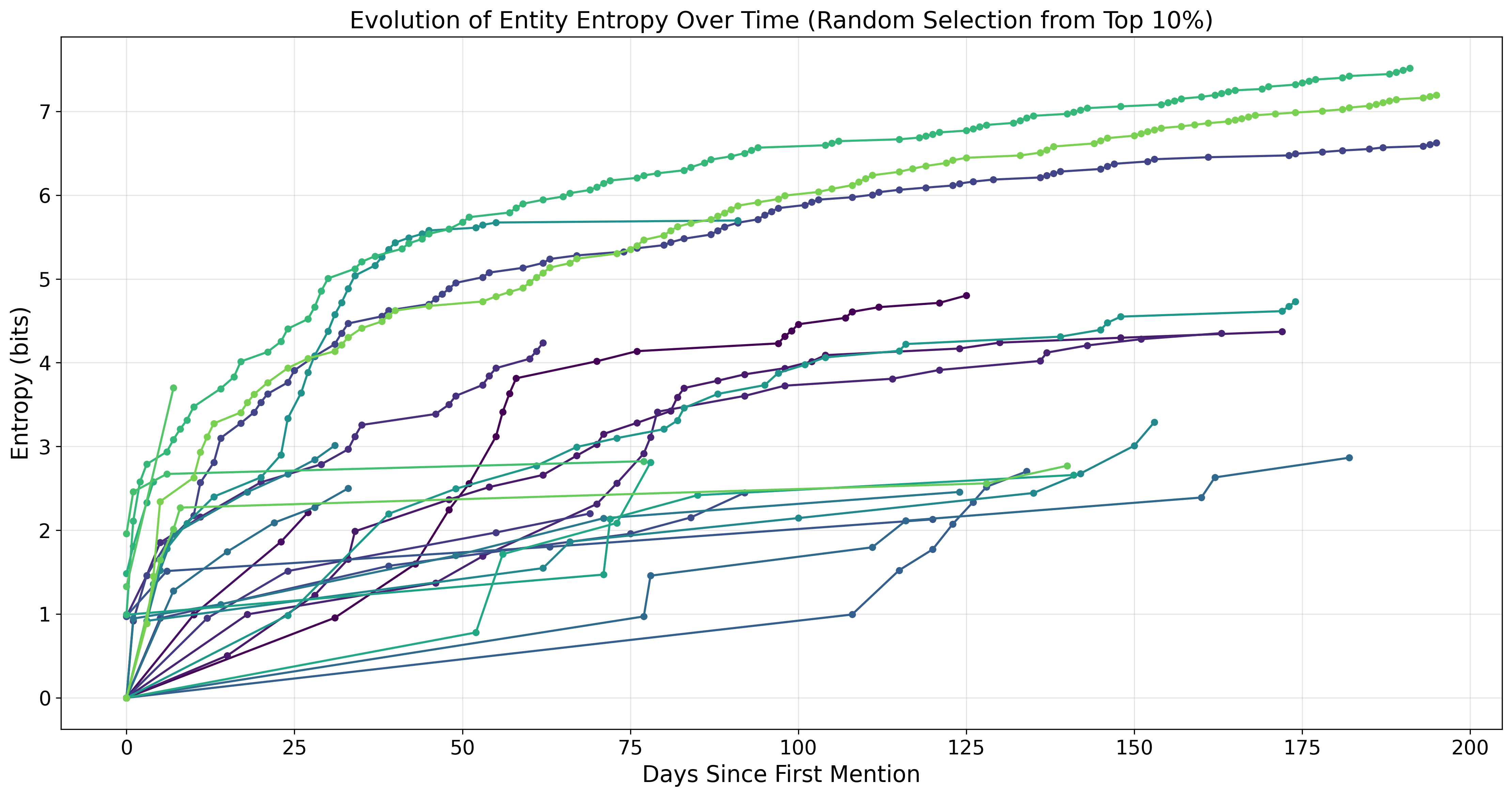}
\caption{Temporal evolution of entropy for 25 randomly selected entities from the top 10\% by current entropy. Note the varied growth patterns over time.}\label{fig:temporal_small}
\end{figure}

On the other hand, we can examine a collection of 25 entities randomly selected from amongst the top 10\% by current entropy. As shown in Figure~\ref{fig:temporal_small}, these entities also exhibit consistently high entropy growth, but with more varied patterns than the top 5. The growth of these entities is more `bursty' and less consistent - this is representative of most entities in our corpus. Although these are relatively high entropy entities, their growth is still varied and at times unpredictable without exogenous context.

The samples above show rapid early growth, slowing as they reach a natural limit in entropy proportional to the number of documents. This leads to a natural hypothesis that early growth may be a signal for an entity's importance to the business.

\begin{figure}[H]
\centering
\includegraphics[width=0.8\textwidth]{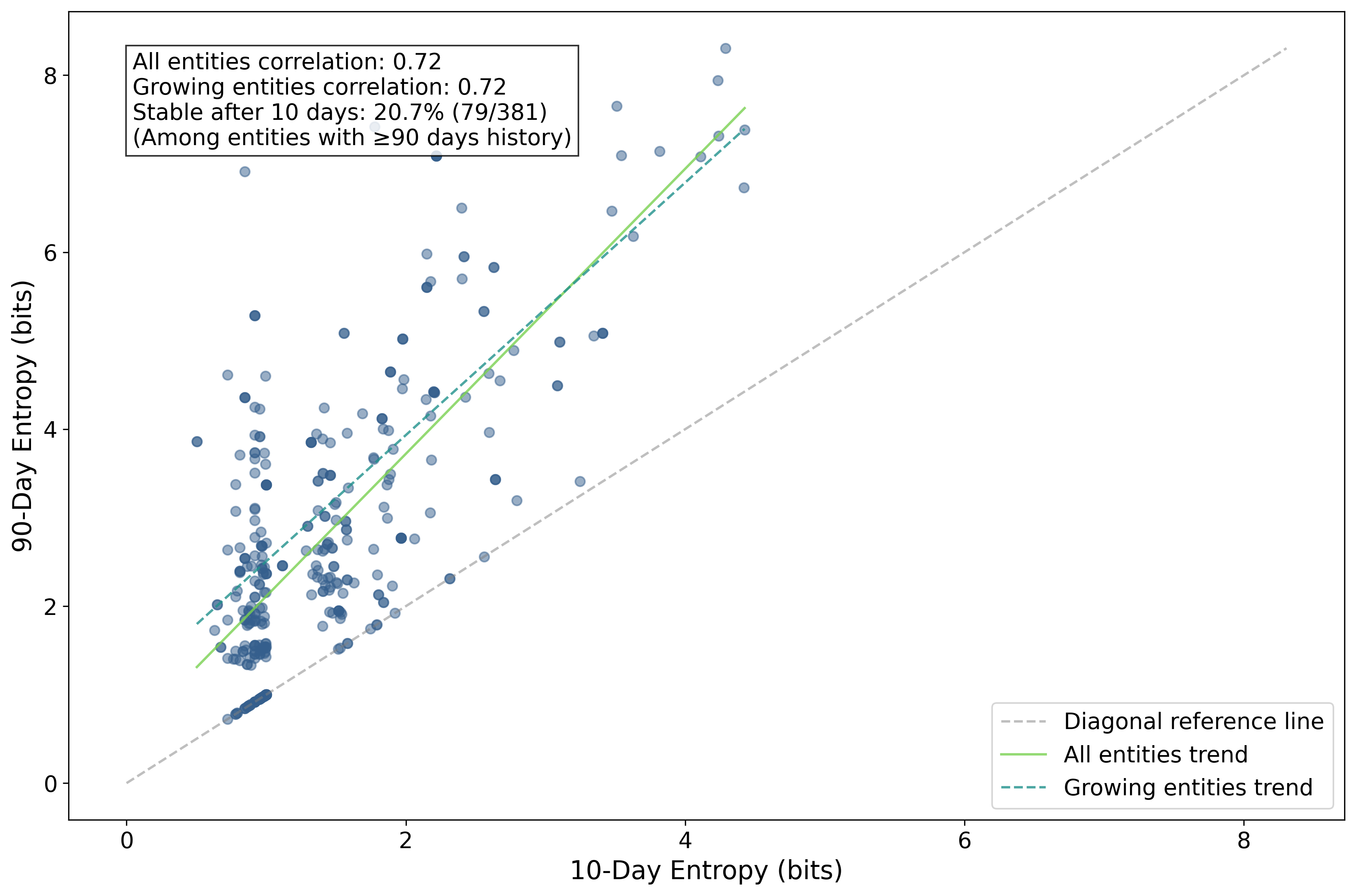}
\caption{Scatter plot showing the relationship between early entropy growth and 90 day entropy for all entities seeing at least one change in entropy. Lighter shaded points indicate there is an overlap of more than one data point. We observe a positive correlation in the 10 day entropy and the 90 day entropy. \textit{Stable Entities} are those that see no increase in entropy and lie on the diagonal (gray). \textit{Growing Entities} are those entities which saw a change in entropy within the first 10 days. We show two trend lines for all entities visualized (solid), and for growing entities (dashed).}\label{fig:early_final}
\end{figure}

It is interesting to note that the first 10 days of an entity's life can be a strong predictor of its final entropy. This is explored in  Figure~\ref{fig:early_final} where we plot the relationship between the 10 day and 90 entropies for entities. Entities that see rapid growth in the first 10 days are more likely to be high entropy entities in the long run. This suggests that early documentation and discussion around an entity are strong indicators of its future complexity and importance to the business.

\subsection{Summary of Empirical Findings}

Our empirical investigation of entity entropy, as detailed in the preceding subsections, reveals a pronounced skew in how knowledge is distributed across the corpus. We highlight the following key observations:

\begin{enumerate}
    \item \textbf{Heavy-Tailed Entropy Distribution (\S\ref{sec:defining_entropic_measures}, \S\ref{sec:empirical_results}):}
    Most entities live within a single document, and a majority of the remainder exhibit low or moderate entropy, indicating that their knowledge is largely concentrated in just a few documents. However, a small minority of entities have significantly higher entropy, wherein facts about them appear across orders of magnitude more documents. This results in a long-tail phenomenon (Figure~\ref{fig:entropy_dist}) that echoes power-law or Pareto-like behaviours \cite{newman2005power}.

    \item \textbf{Relationship Between Entity Size and Entropy (Figure~\ref{fig:size_entropy}):}
    We observe a positive correlation between the total number of facts for an entity (its ``size'') and its entropy. Larger entities not only contain more facts but also disperse them over a larger set of documents. This dispersion grows faster than their mere fact-count expansion, amplifying the overall complexity faced by retrieval systems.

    \item \textbf{Coverage Challenges and Concentration (\S\ref{sec:empirical_results}):}
    High-entropy entities require retrieving many documents for comprehensive coverage (Figure~\ref{fig:coverage}), creating a practical bottleneck for RAG pipelines. By contrast, low-entropy entities -- whose knowledge is captured in one or two documents -- remain straightforward to fetch via standard search and are manageable given current LLM context window size and attention ability.

    \item \textbf{Document Overlap (\S\ref{sec:document_overlap}):}
    Our overlap analysis shows that entities with especially high entropy tend to appear together in multiple documents, forming highly connected hubs in the document-entity graph (Figure~\ref{fig:overlap}). While such overlap can sometimes be beneficial (e.g., a single document covers several related entities), it is not the norm for a majority of the entities.

    \item \textbf{Temporal Patterns and Early Growth (Figures~\ref{fig:temporal}, \ref{fig:temporal_small}, \ref{fig:early_final}):}
    Tracking how entropy evolves from the first mention of an entity reveals that early growth trends are a useful signal for later importance. Entities that see a rapid rise in entropy soon after introduction often end up among the highest-entropy entities in the long run, highlighting the value of early monitoring and intervention (e.g., targeted summarization).

    \item \textbf{Category Differences (Table~\ref{tab:entity-types}):}
    Entities tied to mission-critical or highly visible categories (e.g. core products, the company itself) have higher average entropy. Conversely, straightforward concepts or those confined to niche parts of the organization remain more localized, reinforcing that certain kinds of organizational knowledge inherently resist concentration and may warrant more targeted pre-processing.

\end{enumerate}

Collectively, these findings illustrate how organizational knowledge is unevenly distributed, dominated by a handful of high-entropy entities that drive complexity in search and retrieval. From a practical standpoint, this implies that knowledge management systems can benefit from \emph{selectively} targeting these high-entropy entities for special treatment -- e.g., building dedicated entity-centric summaries or adopting more refined retrieval strategies (Section~\ref{sec:practical_use_cases}). On a theoretical level, recognizing heavy-tailed entropy distributions provides a foundation for modeling the processes by which knowledge fragments over time and across documents, as explored next in Section~\ref{sec:generative_models}.

\FloatBarrier
\section{Generative Models of Entity Knowledge Growth}
\label{sec:generative_models}

Our empirical findings suggest that entity knowledge follows distinct growth patterns, from steady accumulation to sudden bursts of new information. To understand these patterns, we propose a hierarchical entropy-driven generative model that combines baseline importance with entropy feedback at both the time-step and document level in order to model the long term accumulation of entropy. This model is designed to capture the gradual increase in entity entropy over time, while acknowledging the existence of sudden bursts of attention that our model does not yet incorporate.

\subsection{Model Rationale and Objectives}

As observed in Section~\ref{sec:empirical_results}, high-entropy entities exhibit two key behaviours:
\begin{enumerate}
    \item \textbf{Gradual Accumulation of Knowledge}: Many documents are created or updated over time, naturally increasing the entity's coverage.
    \item \textbf{Occasional Bursts}: Periods in which a sudden influx of new documents or facts appears, sharply increasing the entity's entropy in a short time span.
\end{enumerate}

Our generative model focuses on the first behaviour: a steady, cumulative process that leads to increasing entropy. We focus on this as we hypothesize it is related to entropy while these bursts, or ``spikes'', often arise from exogenous factors such as external announcements, strategic reorganizations, or unplanned surges in stakeholder interest. We therefore treat them as out-of-scope for our model while noting that they may be better understood through memetic or diffusion-based approaches in future work \cite{dawkins1976selfish,shifman2013memes}.

\subsection{Hierarchical Model Formulation}
\label{subsec:model_formulation}

In reviewing literature, we looked at diffusion models, such as the Bass model \cite{bass1969new}, which describe the spread of innovations through a population, study of citations and references in academic literature \cite{simkin2007mathematical}, and the spread of information through social networks \cite{shifman2013memes}. The diffusion models we reviewed were limited both through static market sizes as well as binary adoption metrics. Citations are highly related, but offer certain structures not present in our situation (e.g. academic citation reinforces strong hierarchical structures). Social network models are the closest to our situation, but lack the overall coherence and alignment offered in a business context. We therefore developed a novel hierarchical model that captures the gradual accumulation of knowledge about an entity over time using the entropic interpretation introduced above.

Our formulation is inspired by the Bass model \cite{bass1969new}, but extends it to a hierarchical structure that incorporates the entity's historical entropy and allows for document-level variation in the proportion of facts about $E$.

Let $E$ be an entity whose mention footprint evolves over discrete time steps $t = 1, 2, \ldots, T$. Denote by $D_t$ the set of newly created documents at time $t$. For each document $d \in D_t$, the expected number of facts about $E$ is governed by:

\begin{equation}
    \mathbb{E}[f_E(d)] = p_d(E, t) \times \ell_d,
\end{equation}

where:
\begin{itemize}
    \item $p_d(E, t)$ is the document-specific proportion of facts referencing $E$ at time $t$
    \item $\ell_d$ is the total volume of facts in document $d$, drawn from a lognormal distribution:
    \begin{equation}
        \ell_d \sim \mathrm{LogNormal}(\mu,\,\sigma^2)
    \end{equation}
\end{itemize}

The key innovation in our model is the hierarchical structure of $p_d(E, t)$ in which historical entropy helps to guide the time-step mean of $p_d(E, t)$ (particularly over the longterm). Documents created at time $t$ will not have the same proportion of facts, and therefore we model document-level variation through a two-stage process:

\begin{enumerate}
    \item First, compute a time-step mean proportion guided by initial importance and entropy feedback, $\bar{p}(E,t)$:
    \begin{equation}
    \label{eq:p_e_t}
        \bar{p}(E,t) = \sigma\!\Big(\underbrace{\alpha_E\,e^{-\delta_E\,t}}_{\text{baseline importance}}
        + \underbrace{\gamma\big(H_{t-1}(E), H(E)\big)}_{\text{entropy feedback}}\Big)
    \end{equation}

    \item Then, for each document $d$, draw its specific proportion from a Beta distribution centered on $\bar{p}(E,t)$:
    \begin{equation}
        p_d(E,t) \sim \text{Beta}\big(\alpha_{docs}\bar{p}(E,t),\, \alpha_{docs}(1-\bar{p}(E,t))\big)
    \end{equation}
\end{enumerate}

Here:
\begin{itemize}
    \item $\alpha_E$ reflects the entity's initial importance to the organization\footnote{The importance of an entity is a complex measure influenced by exogenous factors and internal organizational dynamics. We treat it as a constant for each entity $E$ in this model but acknowledge that this is a simplistic view.}
    \item $\delta_E > 0$ governs baseline importance over time to capture the natural fading of baseline entity importance (e.g. the importance of the entity given no other factors)
    \item $\gamma(\cdot)$ is a multiplicative function of local and global entropy in order to capture the feedback mechanism caused by entropy. Based on our empirical measurements (Section~\ref{sec:empirical_results}), we hypothesize that highly entropic entity begins to saturate documents and growth slows. On the other hand, if an entity sees a relatively larger entropy during time $t-1$, $H_{t-1}(E)$, than its current overall entropy, $H(E)$, then we would expect this to lead to increased references, increasing the entity's spread across new documents over a short term:
    \begin{equation}
        \gamma(H_{t-1}(E), H(E)) = \frac{1 + \alpha_{local} H_{t-1}(E)}{1 + \alpha_{global}H(E)}
    \end{equation}
    \item $\alpha_{docs}$ controls document-level variation -- higher values lead to tighter clustering around $\bar{p}(E,t)$. Similar to $\alpha_E$, this is a constant for each entity $E$ and is again highly influenced by exogenous factors, for example an Entity introduced through a top-down strategic initiative is likely to have a higher $\alpha_{docs}$ than one introduced within a single team for an operational purpose. Predicting the shape of this distribution is challenging, but we can use the Beta distribution to model this variation by fitting the $\alpha_{docs}$ parameter.
    \item $\sigma(\cdot)$ is the sigmoid function $\sigma(x) = \frac{x}{1+ |x|}$, ensuring proportions remain in $[0,1]$.
\end{itemize}

Each of the alpha parameters are hyperparameters that can be tuned to fit the model to the data and represent the innate `importance' of the entity. We experimented with models in which each were fit using the same parameter, however fits were less flexible and did not capture all entities well. We leave further investigation into the nature of these paramters to future work.

\subsection{Model Examples and Fits}
\label{subsec:model_examples}

With over 3,281 entities in our dataset, a comprehensive display of individual model fits would be impractical. Instead, we present some hand-picked illustrative examples that demonstrate the model's behaviour across different entity types. While these examples represent only a small selection of the overall dataset, they exemplify patterns that consistently emerge across similar entities. We report on summary statistics for all model fits in Appendix \ref{appendix:model_fits}.

\begin{figure}[H]
\centering
\includegraphics[width=0.8\textwidth]{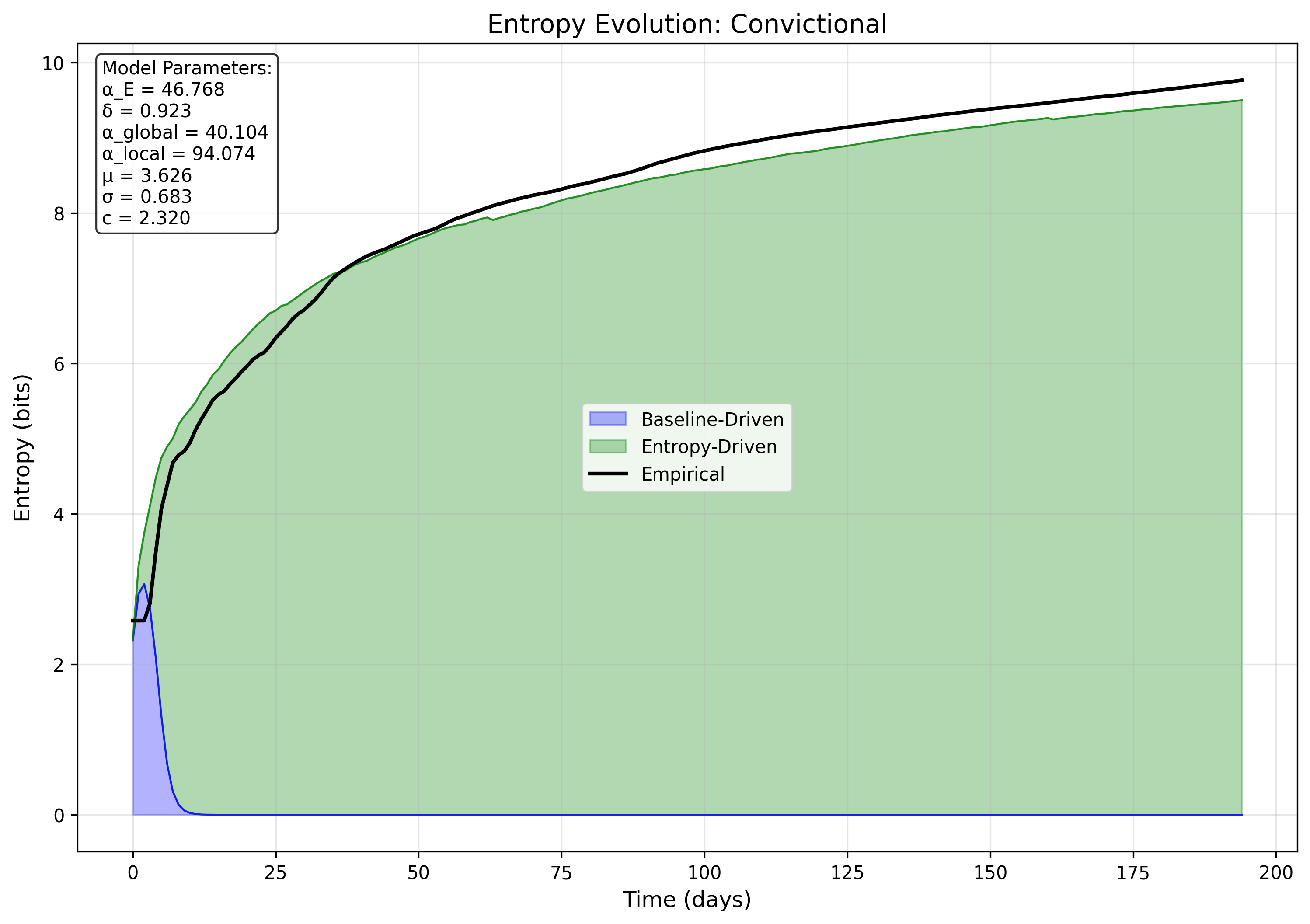}
\caption{Entropy evolution for a high-entropy entity. The solid black line shows empirical data, the blue line and shaded region represent the model's baseline importance contribution, and the green line and shaded region show the model's entropy feedback contribution, adding up to the overall prediction.}\label{fig:convictional}
\end{figure}

Our first example, Figure \ref{fig:convictional}, represents one of the highest-entropy entities in the dataset: the company itself. This entity demonstrates the type of behaviour our model was primarily designed to capture -- steady, long-term growth in entropy as knowledge about the organization accumulates across documents. The model successfully captures this gradual increase, matching both the scale and trajectory of entropy growth over time. However, we observe that the model slightly overestimates entropy during early time steps, particularly during the brief initial period where the entity appears in only a single document. This overestimation suggests that our model might benefit from additional refinement in handling these early-stage transitions.

\begin{figure}[H]
\centering
\includegraphics[width=0.8\textwidth]{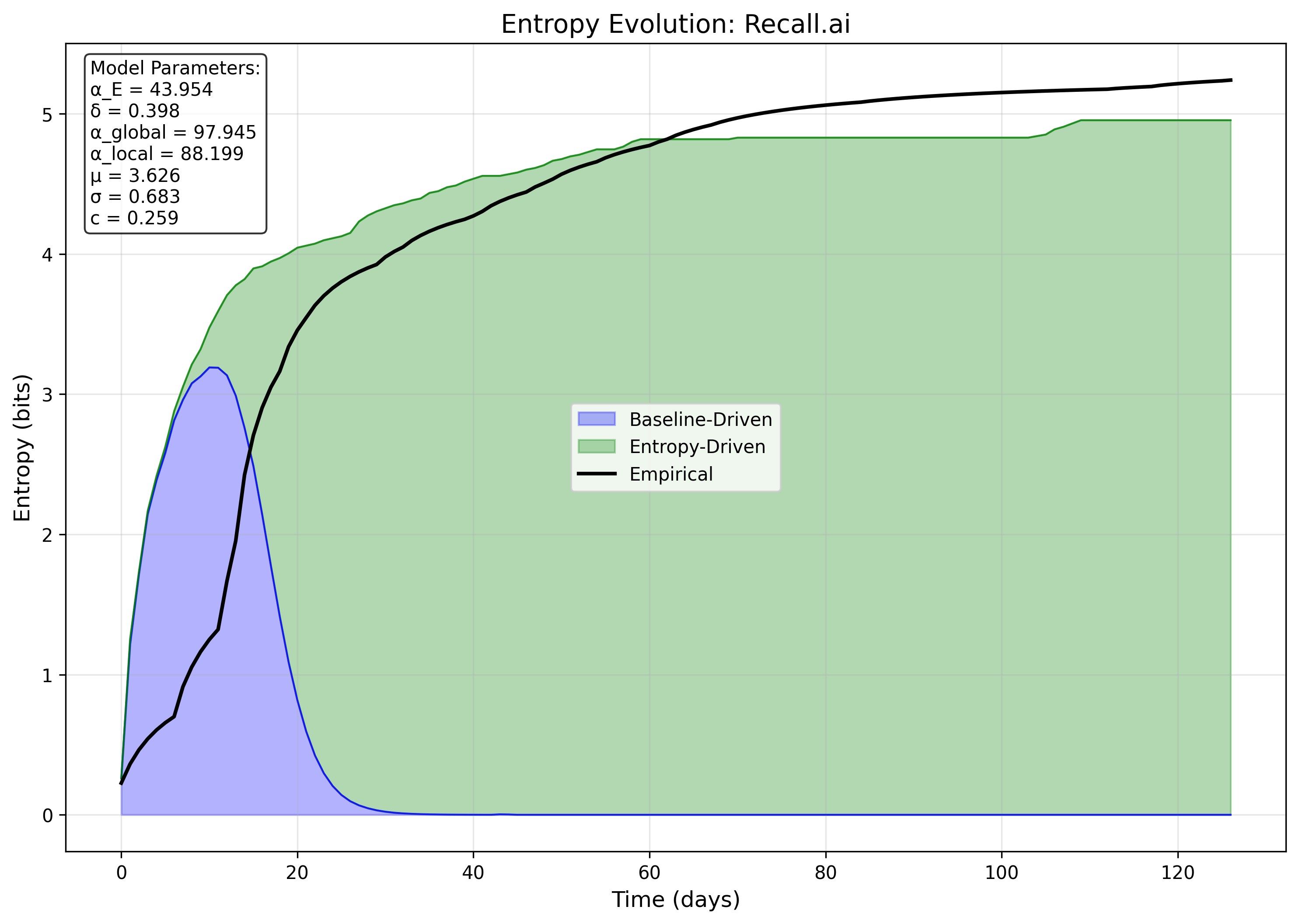}
\caption{Entropy evolution for an entity with bursty growth. Lines and shading follow the same convention as Figure \ref{fig:convictional}.}\label{fig:recall_ai}
\end{figure}

In contrast, Figure \ref{fig:recall_ai} illustrates a case where our model faces significant challenges. This entity, representing an AI tool later integrated into the company's product, exhibits two characteristics that our current model struggles to capture effectively. First, it experiences an extended single-document period where entropy remains at zero. Second, when mentions of the entity do begin to spread across documents, they do so in sudden bursts rather than through gradual accumulation. These bursts of activity, likely triggered by specific organizational events such as evaluation periods or integration decisions, deviate substantially from our model's smooth growth assumptions.

\begin{figure}[H]
\centering
\includegraphics[width=0.8\textwidth]{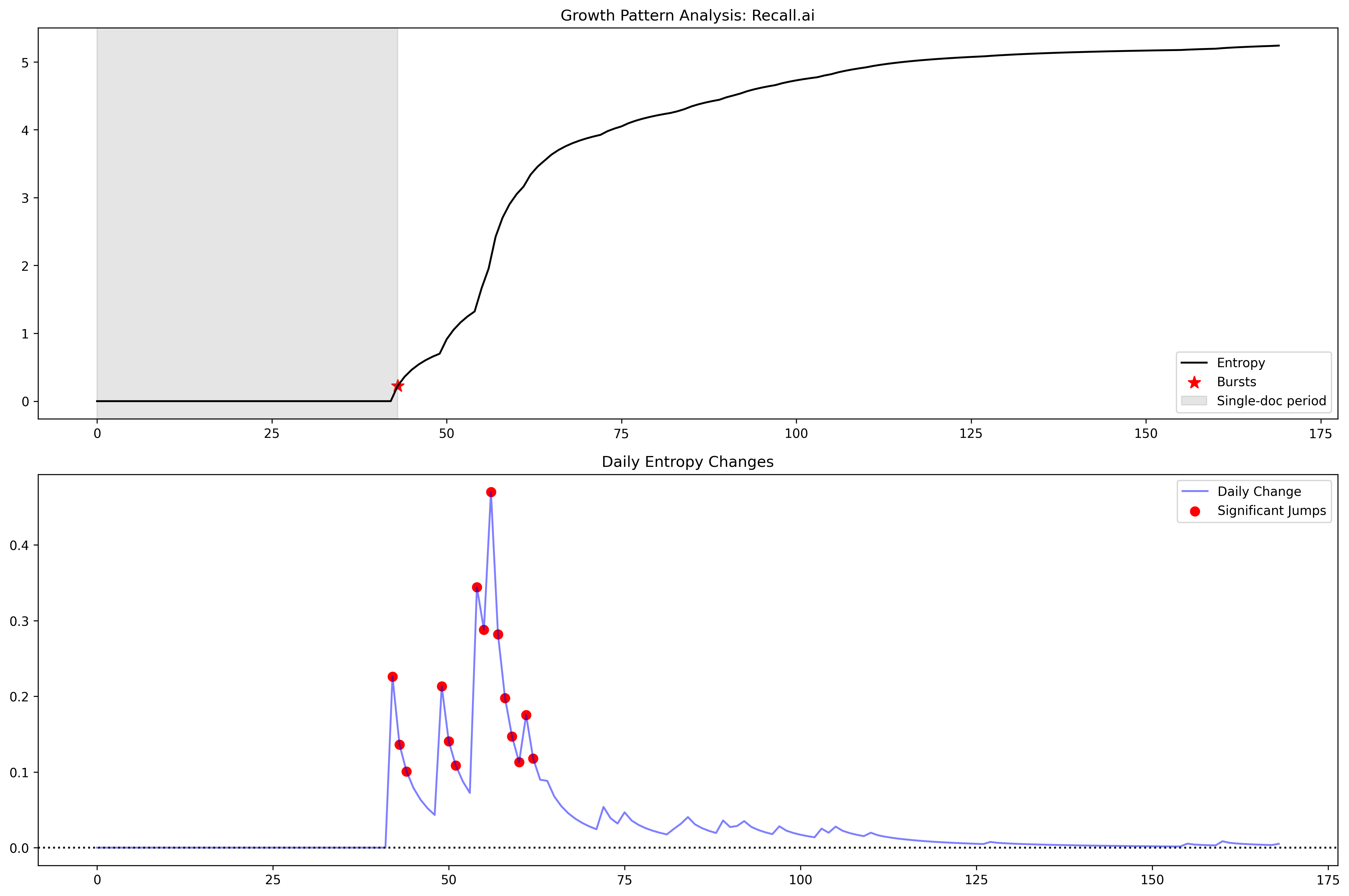}
\caption{Daily entropy changes for the entity shown in Figure \ref{fig:recall_ai}, highlighting discrete jumps in entropy corresponding to organizational events.}\label{fig:bursty}
\end{figure}

To better understand the nature of these bursts, Figure \ref{fig:bursty} provides a detailed view of the daily entropy changes for this same entity. The plot reveals distinct spikes in entropy that correspond to key organizational events: initial discovery of the tool, subsequent evaluation periods, and finally, its selection and integration as a core product feature. This pattern of punctuated growth, while entirely logical from a business perspective, presents a modeling challenge that our current approach does not fully address.

These examples highlight both the strengths and limitations of our current model:

\begin{itemize}
    \item For entities that naturally accumulate mentions over time (like company-wide initiatives or core products), the model provides good fits and meaningful predictions.
    \item However, entities subject to discrete organizational decisions or external events often exhibit ``bursty" behaviour that our smooth growth model cannot adequately capture.
    \item The model's performance during early periods (especially transitions from single to multiple documents) suggests room for improvement in handling these critical phase changes.
\end{itemize}

While our model successfully captures the baseline growth dynamics of entity entropy, these findings point to potential areas for extension, particularly in modeling discrete events and transitions. We discuss possible approaches to addressing these limitations in Section \ref{subsec:potential_extensions}.

\subsection{Document and Fact Creation}
\label{subsec:document_creation}

The model focuses on how entity entropy sprawls within a document corpus; however, the generation of the corpus itself is simply modelled. We observed generally lognormal distributions for document fact counts, as well as documents generated per day. Far from perfect, this provides a strong baseline for the entropy analysis.

One interesting observation is that `bursts' can appear in the document creation process due to outliers, which causes small `wiggles' in the model's growth. This could be indicative of a process for contributing to the bursts where team output is not consistent, but rather comes in waves contributing to the sudden jumps in entropy observed in the data. That said, we hypothesize that the majority of these bursts are exogenous and not simply due to the document creation volume.

\subsection{Potential Extensions of the Model}
\label{subsec:potential_extensions}

Our baseline model produces relatively smooth growth curves reminiscent of logistic or exponential functions over time. Empirically, however, sudden bursts can dramatically alter an entity’s sprawl within days (Figure~\ref{fig:bursty}). These bursts appear to be driven by factors external to an entity’s historical entropy, such as:
\begin{itemize}
    \item Unplanned major announcements (e.g., new funding round).
    \item Organizational shifts in strategy and tactics.
    \item Organizational restructuring or leadership changes.
    \item Viral or memetic events that rapidly propagate across communication channels \cite{shifman2013memes}.
\end{itemize}
Hence, while $p(E,t)$ above captures a \emph{baseline} growth dynamic, we do not account for sudden, exogenous “shocks.” These shocks could be injected as an additional term in Equation~\ref{eq:p_e_t}, or modeled as discrete “burst events” that reset or temporarily boost $p(E,t)$ by a large factor -- both likely adopting a Poisson-like process.

In memetic theory \cite{dawkins1976selfish,shifman2013memes}, cultural artifacts can undergo punctuated growth when they suddenly become a cultural touchstone. Analogously, an organizational concept or project can go “viral” if social or executive interest skyrockets. We see two immediate avenues for improving this baseline generative approach:
\begin{enumerate}
    \item \textbf{Inspiration Shocks:} Extend Equation~\ref{eq:p_e_t} to include an “inspiration” or “burst” factor, $B_t$, that follows a heavy-tailed or point-process distribution \cite{kleinberg2002bursty}. When a burst event happens, $B_t$ spikes and momentarily dominates the mention probability for $E$. However, modelling the timing of these bursts is challenging, as they are often exogenous and unpredictable.
    \item \textbf{Interaction Effects:} Real-world entities often see correlated bursts (e.g., a partner entity might surge in mentions alongside $E$). Modeling cross-entity interactions, such that spikes for one entity propagate to related entities, could further align this approach with memetic or ecological theories of information spread.
\end{enumerate}
Either extension would move closer to capturing the burstiness seen in practice, but it would also introduce additional parameters and complexity. Consequently, one practical approach is to adopt a hybrid strategy: use the baseline model for steady-state updates while implementing a separate detection mechanism (e.g., a threshold on first-derivative of entropy) that flags potential burst events. Once flagged, these events could be handled via specialized data augmentation, entity consolidation, or other retrieval workflows.

\FloatBarrier
\section{Application in RAG Pipelines}
\label{sec:practical_use_cases}

While using entropy as a model for entity sprawl may appear academic, there are practical applications for this model within organizations that can help to improve the efficiency and effectiveness of knowledge retrieval pipelines for LLMs. Advancements in the reasoning abilities of new foundation models (e.g. \texttt{o1 and o1-mini} \cite{openai2025o1}) and substantial increases in the size of the context window have made it possible to retrieve and reason over larger and more complex documents. However, these models are still limited by the amount of information they can process in a single pass. By understanding the entropy of entities within an organization, we can prioritize the pre-processing of complex, high-entropy entities to reduce the amount of information that needs to be processed by the LLM. By using techniques such as fact-extraction and summarization (such as our approach), clustering and summarization (e.g. GraphRAG \cite{edge2024localglobalgraphrag}) or simple LLM summarization \cite{basyal2023textsummarizationusinglarge}, we can reduce the entropy of these entities and make them more completely digestable by an LLM during a single pass. This can lead to more accurate and efficient reasoning over the knowledge contained within an organization by ensuring the LLM has a fulsome view of the entity(s) it is reasoning over.

In order to implement such a pipeline, we suggest employing named entity recognition (even simple key-word) to identify entities within documents during indexing. Once entities are identified, their occurences can be logged allowing for estimation of their entropy. Our analysis shows that the first 10-30 days following an entity's appearence is a strong predictor of its final entropy under this model (although that will vary by organization size, complexity and velocity). By monitoring the growth of entities' entropy over time, we can predict the future complexity of said entities and prioritize the pre-processing of high-entropy entities to improve the efficiency of knowledge retrieval pipelines before degradations in performance become apparent.

\FloatBarrier
\section{Conclusion}
\label{sec:conclusion}

In this paper, we introduced \emph{Business Entity Entropy} as a unifying framework for quantifying how knowledge about an entity is distributed across an organizational corpus. Building on Shannon entropy, we provided a formal measure to capture this ``contextual sprawl,'' showing that many entities remain concentrated in a small number of documents while others exhibit high degrees of dispersion. Through extensive empirical analysis on a large-scale enterprise dataset, we revealed that entity entropy follows a heavy-tailed distribution, where a minority of highly entropic entities drive most of the knowledge fragmentation. These complex entities typically require more sophisticated retrieval and pre-processing strategies to ensure accurate coverage.

Beyond measuring entity entropy, we explored its temporal dynamics and proposed a hierarchical generative model to explain the observed growth patterns. While our baseline model captured only steady incremental gains in entropy, future refinements could incorporate exogenous ``burst'' events that often arise in real-world settings (e.g., strategic shifts, viral product launches, rebranding efforts). We also demonstrated how integrating entity entropy into RAG pipelines helps to prioritize resource-intensive summarization efforts, allowing LLMs to retrieve and reason more effectively about high-entropy entities.

Our findings underscore several directions for future work. First, incorporating temporal shocks and cross-entity interactions could more accurately model real-world bursts of attention. Second, improving fact extraction and de-duplication pipelines would yield richer, more reliable entropy estimates. Third, exploring cross-organizational comparisons of entropy distributions as well as further feature exploration can shed light on whether certain categories or sources  of entities inherently resist knowledge concentration. Finally, from a practical perspective, establishing threshold-based heuristics for early detection of high-entropy entities may streamline knowledge management processes and reduce retrieval overhead in large, evolving corpora.

Overall, we hope that \emph{Business Entity Entropy} proves a useful analytical lens for researchers and practitioners alike, illuminating how organizational knowledge is shaped, shared, and sometimes fragmented. By marrying theoretical insights from information theory with concrete retrieval challenges in enterprise settings, we aim to lay groundwork for more adaptive and robust knowledge retrieval systems in the era of LLMs.

\pagebreak
\FloatBarrier
\section{Appendix} \label{sec:appendix}
\subsection{Fact Extraction Pipeline}
\label{appendix:fact_extraction_pipeline}

Our fact extraction pipeline combines NER and entity-level fact extraction to identify entities and their associated facts. We use a LLM to extract entities from the text and then reverse the process to extract facts from the same document for each identified entity. Here, facts represent individual pieces of information about the entity, such as descriptions, relationships, or timelines. Examples of such might include:

\begin{itemize}
    \item \textsc{Entity:} \textit{Product X}
        \begin{itemize}
        \item \textit{[Fact: Product X is a software tool for data analysis., Confidence: 0.9, Source ID: xxx]}
        \item \textit{[Fact: It was launched in 2019, Confidence: 0.8, Source ID: xyz]}
        \item \textit{[Fact: Since it's launch it has gained popularity among data scientists., Confidence: 0.5, Source ID: xxx]}
        \end{itemize}
    \item \textsc{Entity:} \textit{Person Y}
        \begin{itemize}
        \item \textsc{Facts:} \textit{[Fact: Person Y is the head of the data science team., Confidence: 0.9, Source ID: abc]}
        \item \textit{[Fact: They joined the company in 2017, Confidence: 0.7, Source ID: abc]}
        \item \textit{[Fact: Person Y has a background in machine learning., Confidence: 0.6, Source ID: def]}
        \end{itemize}
    \end{itemize}

The pipeline is designed to be flexible and extensible, allowing for the addition of new entity types and fact categories as needed. The extracted entities and facts are then used to compute the entropy measures as described in Section~\ref{sec:defining_entropic_measures}.

While far from perfect, the pipeline provides a starting point for empirical analysis and theoretical modeling. Future work will focus on improving the accuracy and coverage of the extraction process, as well as exploring more sophisticated entity linking and aggregation techniques including the de-duplication of facts to understand the impact of fact overlap on entity entropy.

\subsection{Model Fits Across All Entities}
\label{appendix:model_fits}

Within Section~\ref{sec:generative_models} about generative models we showed a few examples of how the model performed on a few select entities. In this section we provide a summary of how the model performed across all entities in our dataset with non-zero entropy (e.g. appears in at least two documents).

In total our corpus contained 3,281 entities, of which only 806 exhibited non-zero entropy with at least 180 days of history. We fit the model to each of these entities using \textit{Scipy's} \texttt{L-BFGS-B} optimization algorithm\cite{2020SciPy-NMeth}. The model was fit to each entity individually using 30, 60 and 90 time-steps (days) of the initial non-zero entropy and evaluated on the following 90 days of data. The model was optimized across the following parameters:

For each entity:
\begin{itemize}
    \item $\alpha_E$ - The entity's initial importance to the organization
    \item $\delta_E$ - Baseline importance decay over time
    \item $\alpha_{global}$ - The weight of the historical entropy in the feedback mechanism
    \item $\alpha_{local}$ - The weight of the current entropy in the feedback mechanism
    \item $\alpha_{docs}$ - The document-level variation in the proportion of facts about the entity
\end{itemize}

For all entities:
\begin{itemize}
    \item $\sigma_{facts}$ - The variation in the number of facts per document
    \item $\mu_{facts}$ - The mean number of facts per document
\end{itemize}

We visualize the overall model fits in ‘RMSE’ space in Figure~\ref{fig:fit_metrics}. We plot against the entity's final entropy (in our data) to show how the model better fits entities which have experienced strong entropy feedback. Fitting our data on 30, 60 and 90 days shows the training window becomes more important for entities which showed more consistent growth (resulting in overall high entropy), but is less impactful for entities that do not exhibit high entropy feedback.

Even with 806 non-zero entropy entities in our dataset, when we look for entities with at least 120, 150 and 180 days of history we are left with a decreasing number of entities shown in the plot and summary table below:

\begin{figure}[H]
\centering
\includegraphics[width=0.8\textwidth]{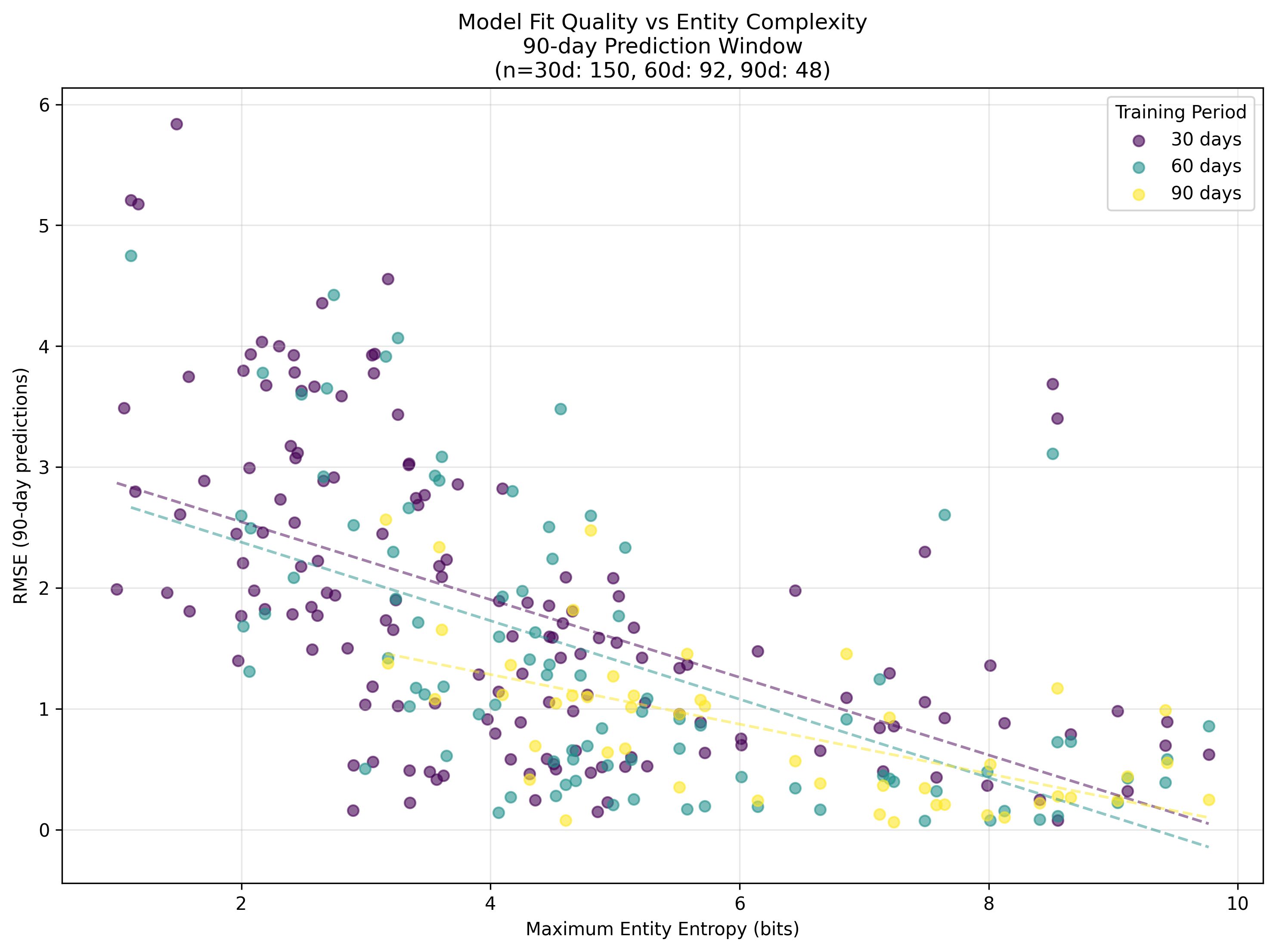}
\caption{Scatter plot showing the RMSE of the model fits across all entities with $> 0$ entropy in the dataset. Each point represents a single entity, with the x-axis showing the total entropy of the entity (across all documents in the corpus). Points are coloured according to the length of training dates used to fit the model.}\label{fig:fit_metrics}
\end{figure}

\begin{table}[H]
\centering
\caption{Model Fit Metrics}
\label{tab:fit_metrics}
\begin{tabular}{|c|c|c|c|c|}
\hline
Training Period (days) & Mean RMSE & Median RMSE & Std RMSE & Valid Samples \\
\hline
30 & 2.2123 & 1.6271 & 2.4427 & 150 \\
60 & 1.8254 & 1.0277 & 2.1796 & 92 \\
90 & 1.0410 & 0.6822 & 1.2276 & 48 \\
\hline
\end{tabular}
\end{table}

\subsection{Fact and Document Distributions}

\begin{figure}[H]
\centering
\includegraphics[width=0.8\textwidth]{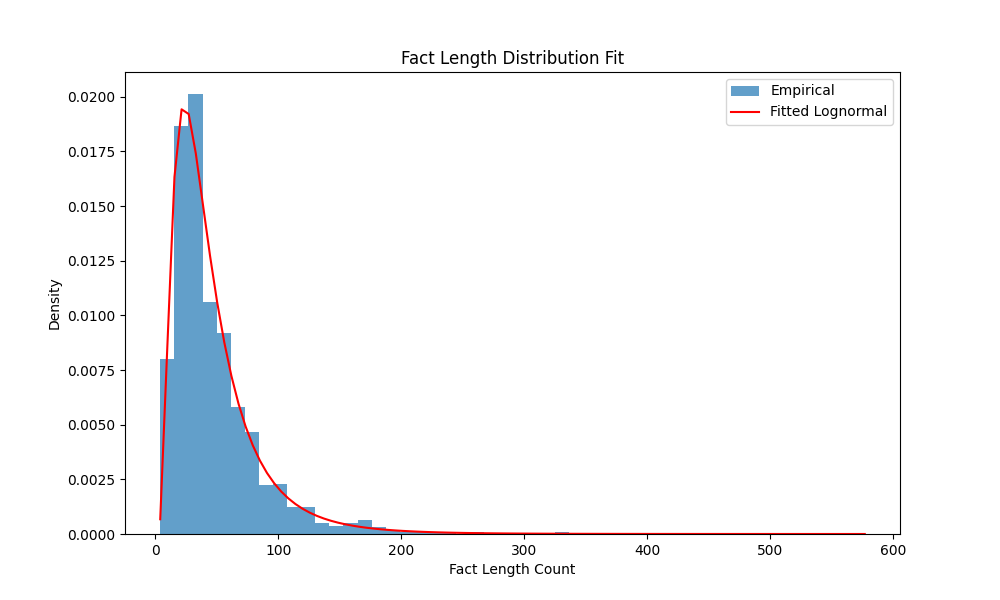}
\caption{Empirical vs. simulated fact distribution across all entities. The empirical distribution is shown in blue, while the simulated distribution is shown in red.}\label{fig:fact_dist_fit}
\end{figure}

Finally, we can visualize the fact distribution fit across all simulated documents against empirical distributions in Figure~\ref{fig:fact_dist_fit}. With over 100,000 individual facts in our empirical data, we were able to observe a fairly smooth lognormal ditribution which allowed for more confident model fits, removing potential error from this process. In isolation we also found this distribution to be both intuitive and interesting, representing a behavioural dynamic in how humans document knowledge.

\subsection{Document Generation}
\label{appendix:document_generation}

\begin{figure}[H]
\centering
\includegraphics[width=0.8\textwidth]{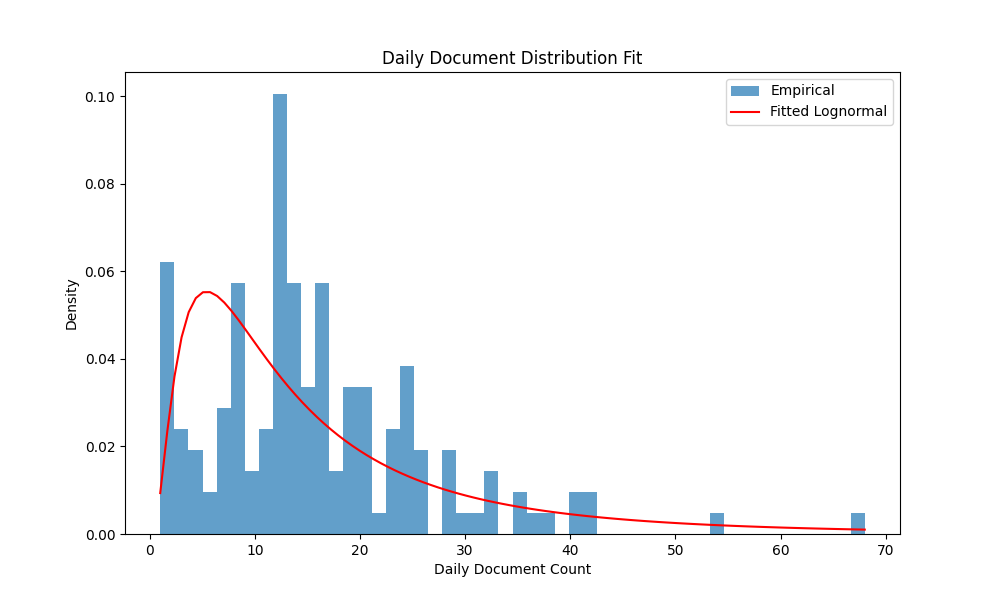}
\caption{Documents generated daily by our organization within Google Drive (documents), Github (Issues and PRs) and objects/activity coming from the Convictional platform.}\label{fig:document_dist_fit}
\end{figure}

In our generative model, we use the actual document counts created daily as opposed to simulating it as we could with facts. This choice was made as the document creation counts did not produce a smooth distribution (see Figure~\ref{fig:document_dist_fit}) and we felt it was important to capture the true nature of the data in order to remove a potential source of error in our model.

\section*{Acknowledgements}
We gratefully acknowledge the support and contributions of the entire team at Convictional, as well as our investors, including Y Combinator and Lachy Groom, and advisors, including Stefano Puntoni and Bart De Langhe. \\[10pt]
We also acknowledge the use of generative AI for help with brainstorming ideas, assisting with analysis code and proof-reading for this paper. \\[10pt]
\noindent Both authors are employees of Convictional, whose mission is to be the infrastructure that powers decisions at the world’s most ambitious companies. For more information about Convictional and our research, please visit \url{https://convictional.com}.

\pagebreak
\bibliographystyle{plain}
\bibliography{references}

\begin{thebibliography}{10}

\bibitem{bass1969new}
Frank~M. Bass.
\newblock A new product growth model for consumer durables.
\newblock {\em Management Science}, 15(5):215--227, 1969.

\bibitem{basyal2023textsummarizationusinglarge}
Lochan Basyal and Mihir Sanghvi.
\newblock Text summarization using large language models: A comparative study
  of mpt-7b-instruct, falcon-7b-instruct, and openai chat-gpt models, 2023.

\bibitem{chieu2003named}
Hai~Leong Chieu and Hwee~Tou Ng.
\newblock Named entity recognition with a maximum entropy approach.
\newblock In {\em Proceedings of the Seventh Conference on Natural Language
  Learning at HLT-NAACL 2003}, pages 160--163. Association for Computational
  Linguistics, 2003.

\bibitem{cover1991elements}
Thomas~M Cover and Joy~A Thomas.
\newblock Elements of information theory.
\newblock {\em John Wiley \& Sons}, 1991.

\bibitem{dawkins1976selfish}
Richard Dawkins.
\newblock {\em The Selfish Gene}.
\newblock Oxford University Press, 1976.

\bibitem{edge2024localglobalgraphrag}
Darren Edge, Ha~Trinh, Newman Cheng, Joshua Bradley, Alex Chao, Apurva Mody,
  Steven Truitt, and Jonathan Larson.
\newblock From local to global: A graph rag approach to query-focused
  summarization, 2024.

\bibitem{kleinberg2002bursty}
Jon~M. Kleinberg.
\newblock Bursty and hierarchical structure in streams.
\newblock In {\em Proceedings of the Eighth ACM SIGKDD International Conference
  on Knowledge Discovery and Data Mining (KDD)}, pages 91--101. ACM, 2002.

\bibitem{lewis2020retrieval}
Patrick Lewis, Ethan Perez, Aleksandra Piktus, Fabio Petroni, Vladimir
  Karpukhin, Naman Goyal, Heinrich Küttler, Mike Lewis, Wen-tau Yih, Tim
  Rocktäschel, et~al.
\newblock Retrieval-augmented generation for knowledge-intensive nlp tasks.
\newblock {\em Advances in Neural Information Processing Systems}, 33, 2020.

\bibitem{newman2005power}
MEJ Newman.
\newblock Power laws, pareto distributions and zipf’s law.
\newblock {\em Contemporary Physics}, 46(5):323–351, September 2005.

\bibitem{nonaka1994dynamic}
Ikujiro Nonaka.
\newblock A dynamic theory of organizational knowledge creation.
\newblock {\em Organization Science}, 5(1):14--37, 1994.

\bibitem{openai2025o1}
Jakub Pachocki, Jerry Tworek, Liam Fedus, Lukasz Kaiser, Mark Chen, Szymon
  Sidor, and Wojciech Zaremba.
\newblock Learning to reason with {LLMs}.
\newblock Blog post, OpenAI, 2025.

\bibitem{ratinov2009design}
Lev Ratinov and Dan Roth.
\newblock Design challenges and misconceptions in named entity recognition.
\newblock In {\em Proceedings of the Thirteenth Conference on Computational
  Natural Language Learning (CoNLL-2009)}, pages 147--155. Association for
  Computational Linguistics, 2009.

\bibitem{shannon1948mathematical}
Claude~E Shannon.
\newblock A mathematical theory of communication.
\newblock {\em Bell System Technical Journal}, 27(3):379--423, 1948.

\bibitem{shifman2013memes}
Limor Shifman.
\newblock {\em Memes in Digital Culture}.
\newblock MIT Press, 2013.

\bibitem{simkin2007mathematical}
Mikhail~V Simkin and Vwani~P Roychowdhury.
\newblock A mathematical theory of citing.
\newblock {\em Journal of the American Society for Information Science and
  Technology}, 58(11):1661--1673, 2007.

\bibitem{2020SciPy-NMeth}
Pauli Virtanen, Ralf Gommers, Travis~E. Oliphant, Matt Haberland, Tyler Reddy,
  David Cournapeau, Evgeni Burovski, Pearu Peterson, Warren Weckesser, Jonathan
  Bright, St{\'e}fan~J. {van der Walt}, Matthew Brett, Joshua Wilson, K.~Jarrod
  Millman, Nikolay Mayorov, Andrew R.~J. Nelson, Eric Jones, Robert Kern, Eric
  Larson, C~J Carey, {\.I}lhan Polat, Yu~Feng, Eric~W. Moore, Jake
  {VanderPlas}, Denis Laxalde, Josef Perktold, Robert Cimrman, Ian Henriksen,
  E.~A. Quintero, Charles~R. Harris, Anne~M. Archibald, Ant{\^o}nio~H. Ribeiro,
  Fabian Pedregosa, Paul {van Mulbregt}, and {SciPy 1.0 Contributors}.
\newblock {{SciPy} 1.0: Fundamental Algorithms for Scientific Computing in
  Python}.
\newblock {\em Nature Methods}, 17:261--272, 2020.

\end{thebibliography}

\phantomsection\addcontentsline{toc}{section}{References}

\end{document}